\documentclass[final,leqno,oneside,notitlepage,10pt,twocolumn]{article}

\usepackage{amsthm}
\usepackage{thmtools}
\usepackage{graphicx}
\usepackage[margin=2cm,font=normalsize,labelfont=bf]{caption}
\usepackage{verbatim}
\usepackage[shortlabels]{enumitem}
\usepackage{fancyhdr}
\usepackage[]{geometry}
\usepackage{xstring}
\usepackage{needspace}
\usepackage{color}
\usepackage{amstext}
\usepackage{nameref}
\usepackage[hypertexnames=false]{hyperref}
\usepackage{mathdots}
\usepackage{soul}
\usepackage{chngcntr}
\usepackage[section]{placeins}

\makeatletter
\newcounter{denseversion}
\newcounter{comments}
\newcounter{authorcounter}
\newcounter{adresscounter}

\def\title#1{\gdef\@title{#1}}
\def\@title{}
\def\subtitle#1{\gdef\@subtitle{#1}}
\def\@subtitle{}

\def\authortagsused{0}
\def\adresstag#1{\if!#1!\else$^{\;#1\;}$\fi}
\def\@authorsep#1{
  \ifnum\value{authorcounter}=#1 and \else\unskip, \fi
}

\renewcommand{\author}[2][]{
  \stepcounter{authorcounter}
  \if!#1!\else\gdef\authortagsused{1}\fi
  \ifnum\value{authorcounter}=1
    \def\@authorstringa{#2\adresstag{#1}}
    \def\@authorstringb{#2}
    \def\@authorstringc{#2\adresstag{#1}}
  \else
    \ifnum\value{authorcounter}=2
      \g@addto@macro\@authorstringa{\@authorsep{2}#2\adresstag{#1}}
      \g@addto@macro\@authorstringb{\@authorsep{2}#2}
      \g@addto@macro\@authorstringc{\\#2\adresstag{#1}}
    \else
      \g@addto@macro\@authorstringa{\@authorsep{3}#2\adresstag{#1}}
      \g@addto@macro\@authorstringb{\@authorsep{3}#2}
      \g@addto@macro\@authorstringc{\\#2\adresstag{#1}}
    \fi
  \fi}
\def\@author{\ifnum\value{denseversion}=0\@authorstringa\else\@authorstringb\fi}

\def\@adressstringa{}
\def\@adressstringb{}
\newcommand{\adress}[2][]{
  \stepcounter{adresscounter}
  \ifnum\value{adresscounter}=1
    \g@addto@macro\@adressstringa{\ifnum\authortagsused=0\def\br{\\}\else\def\br{, }\fi\adresstag{#1}#2}
    \g@addto@macro\@adressstringb{\def\br{\\}\adresstag{#1}\parbox[t]{14cm}{#2}}
  \else
    \g@addto@macro\@adressstringa{\\[\bigskipamount]\adresstag{#1}#2}
    \g@addto@macro\@adressstringb{\\[\medskipamount]\adresstag{#1}\parbox[t]{14cm}{#2}}
  \fi}

\def\preprint#1{\gdef\@preprint{#1}}
\def\@preprint{}
\def\keywords#1{\gdef\@keywords{#1}}
\def\@keywords{}
\def\msc#1{\gdef\@msc{#1}}
\def\@msc{}
\def\email#1{
   \gdef\@email{#1}
   \g@addto@macro\@authorstringc{ {\it (#1)}}}
\def\@email{}
\def\dedication#1{\gdef\@dedication{#1}}
\def\@dedication{}

\def\mybaselinestretch#1{
  \gdef\@mybaselinestretch{#1}
  \renewcommand{\baselinestretch}{\@mybaselinestretch}}

\def\myparskip#1{
  \gdef\@myparskip{#1}
  \setlength{\parskip}{\@myparskip}}

\mybaselinestretch{1.2}
\myparskip{1.5ex}

\binoppenalty=\maxdimen
\relpenalty=\maxdimen

\normalfont
\normalsize
\newlength{\@listleftmargin}
\def\setenumstandard{
  \setlist{leftmargin=\@listleftmargin,itemsep=0pt,topsep=0pt,partopsep=0pt,parsep=\@myparskip}
  \setlist[enumerate]{align=left,labelsep=*,leftmargin=\@listleftmargin,itemsep=0pt,topsep=0pt,partopsep=0pt,parsep=\@myparskip}
}

\def\denseversion{
  \setcounter{denseversion}{1}
  \newgeometry{left=3cm,right=3cm,top=3cm}
  \mybaselinestretch{1.1}
  \myparskip{0.8ex}
  \normalfont
  \def\possiblelinebreak{}
  \fancyfoot[C]{\itshape{--$\,\,$\thepage$\,\,$--}}}

\def\possiblelinebreak{\\}

\renewcommand{\emph}[1]{\def\reserved@a{it}\ifx\f@shape\reserved@a\ul{#1}\else\textit{#1}\fi}

\def\setcrefnames{}
\newcommand{\mytableofcontents}{
   \ifnum\value{denseversion}=0
     \tableofcontents
     \setcrefnames 
   \else
     \renewcommand{\baselinestretch}{1.1}
     \setlength{\parskip}{0ex}
     \normalfont
     \begingroup
     \def\addvspace##1{\vskip0.4em}
     \tableofcontents
     \setcrefnames 
     \endgroup
     \renewcommand{\baselinestretch}{\@mybaselinestretch}
     \setlength{\parskip}{\@myparskip}
     \normalfont
   \fi}

\newlength{\zeilenlaenge}
\def\putindent#1{
  \settowidth{\zeilenlaenge}{#1}
  \ifnum\zeilenlaenge>\textwidth
    #1
  \else
    \noindent #1
  \fi
}

\hypersetup{
    linktocpage = true,
    colorlinks,
    linkcolor=[RGB]{0,0,120},
    citecolor=[RGB]{0,0,120},
    urlcolor=[RGB]{0,0,120}
}

\def\pdfdaten{
  \hypersetup{
    pdftitle = {\@title},
    pdfauthor = {\@author},
    pdfkeywords = {\@keywords},    
    bookmarksopen = true,
    bookmarksopenlevel = 1
  }}  

\def\showkeywords{\begin{flushleft}\footnotesize\textbf{Keywords}: \@keywords\end{flushleft}}
\def\showmsc{\begin{flushleft}\footnotesize\textbf{MSC 2010}: \@msc\end{flushleft}}

\def\tocsection#1{\section*{#1}\addcontentsline{toc}{section}{#1}}

\def\mytitle{}
\def\zmptitle{
  \begin{tabular}{cc}
    \begin{minipage}[c]{0.4\textwidth}
      \begin{flushleft}
        \includegraphics[width=110pt]{../../tex/zmp}
      \end{flushleft}  
    \end{minipage}&
    \begin{minipage}[c]{0.55\textwidth}
      \begin{flushright}
      {\small\sf\@preprint}
      \end{flushright}
    \end{minipage}
  \end{tabular}
  \vskip 2cm}

\def\maketitle{
  \pdfdaten
  \noindent
  \mytitle
  \begin{center}
    \LARGE\@title\\
    \if!\@subtitle!\else\smallskip\LARGE\@subtitle\\\fi
    \bigskip
    \if!\@author!\else\bigskip\large\@author\\\fi
    \ifnum\value{denseversion}=0
      \if!\@adressstringa!\else\bigskip\normalsize\@adressstringa\\\fi
      \if!\@email!\else\ifnum\value{authorcounter}=1\bigskip\normalsize\textit{\@email}\\\else\fi\fi
    \else
    \fi
    \if!\@dedication!\else\bigskip\normalsize{\@dedication}\\\fi
  \end{center}
  \ifnum\value{denseversion}=0\vskip 1.5cm\else\vskip0.5cm\fi}

\def\kobib#1{
  \begin{raggedright}
  \ifnum\value{denseversion}=0\else\small\fi
  \Oldbibliography{#1/kobib}
  \bibliographystyle{#1/kobib}
  \end{raggedright}
  \ifnum\value{denseversion}=0\else
      \noindent
      \if!\@authorstringc!\else
        \ifnum\authortagsused=0\ifnum\value{authorcounter}>1\normalsize\@authorstringc\\[\medskipamount]\else\fi\else\normalsize\@authorstringc\\[\medskipamount]\fi
      \fi
      \if!\@adressstringb!\else\normalsize\@adressstringb\\{}\fi
      \ifnum\authortagsused=0
        \ifnum\value{authorcounter}=1
          \if!\@email!\else\linebreak\normalsize\textit{\@email}\\{}\fi
        \else
        \fi
      \else
      \fi
  \fi}

\let\Oldbibliography\bibliography
\def\bibliography#1{
  \begin{raggedright}
  \ifnum\value{denseversion}=0\else\small\fi
  \Oldbibliography{#1}
  \end{raggedright}
  \ifnum\value{denseversion}=0\else
      \medskip
      \noindent
      \if!\@authorstringc!\else
        \ifnum\authortagsused=0\ifnum\value{authorcounter}>1\normalsize\@authorstringc\\[\medskipamount]\else\fi\else\normalsize\@authorstringc\\[\medskipamount]\fi
      \fi
      \if!\@adressstringb!\else\normalsize\@adressstringb\\{}\fi
      \ifnum\authortagsused=0
        \ifnum\value{authorcounter}=1
          \if!\@email!\else\linebreak\normalsize\textit{\@email}\\{}\fi
        \else
        \fi
      \else
      \fi
  \fi
}

\newenvironment{commentfigure}{}
\newenvironment{sidewayscommentfigure}{\begin{minipage}}{\end{minipage}}
\newenvironment{displaycomment}{\begin{list}{}{\rightmargin=1cm\leftmargin=1cm}\item\sf\begin{small}}{\end{small}\end{list}}

\def\tocmark#1{}

\def\draftstamp#1{
  \def\tocmark##1{
    \ifnum\c@secnumdepth=0\section{##1}\fi
    \ifnum\c@secnumdepth=1\subsection{##1}\fi
    \ifnum\c@secnumdepth=2\subsubsection{##1}\fi
    \ifnum\c@secnumdepth=3\subsubsection{##1}\fi
  }
  \ifnum\value{comments}=0
    \gdef\@draft{DRAFT - Edited on \today\ by #1 - Comments are not displayed}
  \else
    \gdef\@draft{DRAFT - Edited on \today\ by #1 - Comments are displayed}
  \fi
  \fancyhead[C]{\footnotesize\tt\textcolor{red}{\@draft}}}

\def\skript{
  \renewenvironment{displaycomment}{}{}
  \ifnum\value{comments}=0
    \renewenvironment{example*}{\comment}{\endcomment}
    \renewenvironment{remark*}{\comment}{\endcomment}
  \else\fi
  \parindent=0mm	
}

\newcounter{sectioning}
\def\tsection#1{\ifnum\value{sectioning}=0\section{#1}\fi}
\def\lsection#1{
  \ifnum\value{sectioning}=1
    \clearpage
    \def\thesection{\lektionname~\arabic{section}:}
    \section{#1}
    \def\thesection{\arabic{section}}
  \fi}
\def\tsubsection#1{\ifnum\value{sectioning}=0\subsection{#1}\fi}
\def\lsubsection#1{\ifnum\value{sectioning}=1\subsection{#1}\fi}
\def\tsubsubsection#1{\ifnum\value{sectioning}=0\subsubsection{#1}\fi}
\def\lsubsubsection#1{\ifnum\value{sectioning}=1\subsubsection{#1}\fi}

\mybaselinestretch{}
\setenumstandard

\makeatother

\usepackage{amssymb}
\usepackage{amsmath}
\usepackage{amstext}
\usepackage{mathrsfs}
\usepackage{euscript}
\usepackage{color}
\usepackage[all,cmtip,color,arrow]{xy}
\usepackage{xstring}
\usepackage{slashed}
\usepackage{tensor}
\usepackage{tikz}

\entrymodifiers={+!!<0pt,\fontdimen22\textfont2>}

\def\Z {\mathbb{Z}}

\def\R {\mathbb{R}}
\def\C {\mathbb{C}}

\def\im{\mathrm{i}}
\def\id{\mathrm{id}}

\def\hc#1{\mathrm{h}_{#1}}
\def\h {\mathrm{H}}

\def\subset{\subseteq}

\renewcommand{\varepsilon}{\epsilon}



\def\notebox#1#2{\begin{minipage}[b]{#1}\sloppy\renewcommand{\baselinestretch}{0.8}\footnotesize \begin{center}#2\end{center}\end{minipage}}

\newcommand{\arr}[1][r]{\ar@<0.7ex>[#1]\ar@<-0.7ex>[#1]}
\newcommand{\arrr}[1][r]{\ar@<1.4ex>[#1]\ar[#1]\ar@<-1.4ex>[#1]}
\newcommand{\arrrr}[1][r]{\ar@<2.1ex>[#1]\ar@<-2.1ex>[#1]\ar@<0.7ex>[#1]\ar@<-0.7ex>[#1]}

\def\stackref#1#2{\stackrel{\text{\ref{#1}}}{#2}}
\def\eqref#1{\stackref{#1}{=}}

\newlength{\myeqt} 
\newlength{\myeqs} 
\newlength{\myeqm} 
\newlength{\myeqn} 
\newcommand\symtext[3][\myeqn]{
  \settowidth{\myeqt}{#2}
  \settowidth{\myeqs}{$#3$}
  \addtolength{\myeqs}{\the\myeqm}
  \ifdim\myeqt>\myeqs
    \stackrel{\hspace{-#1}\notebox{#1}{\medskip #2 \\ $\downarrow$\smallskip}\hspace{-#1}}{#3}
  \else
    \stackrel{\text{#2}}{#3}
  \fi}

\def\brackets#1{\IfStrEq{#1}{-}{}{(#1)}}
\def\subindex#1{\IfStrEq{#1}{-}{}{_{#1}}}

\newcommand{\alxydim}[2]{\begin{aligned}\xymatrix#1{#2}\end{aligned}}



\newlength{\myl}

\def\ddt#1#2#3{\left.\frac{\mathrm{d}^{\IfStrEq{#1}{1}{}{#1}}}{\mathrm{d}#2}\IfStrEq{#2}{#3}{\right.}{\right|_{#3}}}

\def\aut{\mathrm{Aut}}
\def\AUT{\mathrm{AUT}}

\newcommand{\ueins}{{\mathrm{U}}(1)}
\newcommand{\ug}[1]{{\mathrm{U}}\brackets{#1}}

\newcommand{\gl}[1]{{\mathrm{GL}}\brackets{#1}}
\newcommand{\su}[1]{{\mathrm{SU}}\brackets{#1}}

\newcommand{\spin}[1]{{\mathrm{Spin}}\brackets{#1}}
\def\Spin{\spin}
\newcommand{\spinc}[1]{{{\mathrm{Spin}}^{{c}}}\brackets{#1}}

\def\so#1{{\mathrm{SO}}\brackets{#1}}
\renewcommand{\o}[1]{{\mathrm{O}}\brackets{#1}}

\newcommand{\str}[1]{{\mathrm{String}}\brackets{#1}}

\def\ev{\mathrm{ev}}

\def\pr{{\mathrm{pr}}}

\newlength{\widthtmp}
\def\length#1{\settowidth{\widthtmp}{#1}\the\widthtmp}

\def\ubun#1{\bun\relax{#1}}

\def\ubuncon#1{\buncon\relax{#1}}
\def\ubunconflat#1{\bunconflat\relax{#1}}

\def\ufusbun#1{\mathcal{F}\!us\buntech{}{}(#1)}

\def\ugrb#1{\grb{\,}{#1}}
\def\ugrbcon#1{\grbcon\relax{#1}}

\usepackage[latin1]{inputenc}
\usepackage[english]{babel}
\usepackage[english]{cleveref}

\def\refname{References}

\def\quot#1{``#1''}

\def\nameTheorem{Theorem}
\def\nameTheorems{Theorems}
\def\nameDefinition{Definition}
\def\nameDefinitions{Definitions}
\def\nameCorollary{Corollary}
\def\nameCorollaries{Corollaries}
\def\nameProposition{Proposition}
\def\namePropositions{Propositions}
\def\nameConstruction{Construction}
\def\nameConstructions{Constructions}
\def\nameLemma{Lemma}
\def\nameLemmas{Lemmas}
\def\nameRemark{Remark}
\def\nameRemarks{Remarks}
\def\nameProblem{Problem}
\def\nameProblems{Problems}
\def\nameExample{Example}
\def\nameExamples{Examples}
\def\nameExercise{Exercise}
\def\nameExercises{Exercises}

\def\nameProof{Proof}
\def\nameEquation{Eq.}
\def\nameEquations{Eqs.}
\def\nameSection{Section}
\def\nameSections{Sections}
\def\nameAppendix{Appendix}
\def\nameAppendixs{Appendices}
\def\nameFigure{Figure}
\def\nameFigures{Figures}

\input{theorems}
\input{kobib}

\title{String structures and loop spaces}
\author{Konrad Waldorf}
\email{konrad.waldorf@uni-greifswald.de}

\adress{
  Universität Greifswald\\
  Institut für Mathematik und Informatik\\
  Walther-Rathenau-Str. 47\\
  D-17487 Greifswald}

\msc{}
\dedication{Contribution to the Encyclopedia of Mathematical Physics (2nd edition)}
\pdfdaten

\denseversion
\nolinks
\allowdisplaybreaks

\usepackage{amstext}

\def\mathscr#1{\EuScript{#1}}
\newcommand{\XString}{\mathscr{S}\mathrm{tring}}
\def\AUT{\mathscr{A}\mathrm{ut}}

\def\pfaff{\mathrm{pf}}
\def\Pfaff{\mathrm{Pf}}

\def\Iso{\mathscr{I}\mathrm{so}}
\def\ugrb#1{\mathscr{G}\mathrm{rb}\brackets{#1}}
\def\ugrbcon#1{\mathscr{G}\mathrm{rb}^{\nabla}\brackets{#1}}
\def\ubun#1{\mathscr{LB}\mathrm{dl}\brackets{#1}}
\def\ufusbun#1{\mathscr{F}\mathrm{us}\mathscr{LB}\mathrm{dl}\brackets{#1}}
\def\ubuncon#1{\mathscr{LB}\mathrm{dl}^{\nabla}\brackets{#1}}

\def\ubunconflat#1{\mathscr{LB}\mathrm{dl}^{\nabla_0}\brackets{#1}}
\def\vN{N_{\mathrm{III}_1}}

\creflabelformat{section}{#2\S#1#3}
\crefname{section}{\unskip}{\unskip}
\crefname{equation}{\unskip}{\unskip}

\usepackage{natbib}
\setcitestyle{authoryear,open={(},close={)}}

\mybaselinestretch{1.0165}
\myparskip{0.5ex}

\usepackage{graphicx}

\begin{document}

\twocolumn[\maketitle] 

\begin{abstract}
We provide a concise and accessible introduction to (geometric) string structures, highlighting their connection to loop spaces and outlining relationships with neighboring topics.
\end{abstract}

\mytableofcontents

\setsecnumdepth{1}

\section{Introduction}

\label{introduction}

 String structures are tangential structures on a smooth manifold. In general, tangential structures alter the structure group of the frame bundle of a smooth manifold $M$ through a homomorphism
\begin{equation*}
\rho: G \to \gl d
\end{equation*}
 of topological groups, where $d$ is the dimension of $M$ and $\mathrm{GL}(d)$ is the general linear group of $\mathbb{R}^d$. In terms of principal bundles, a $G$-structure on  $M$ is a principal $G$-bundle $P$ together with a continuous map $P \to \mathrm{GL}(M)$ to the frame bundle of $M$  that commutes with the projections to $M$ and is $G$-equivariant under the $G$-action on $\mathrm{GL}(M)$ induced by $\rho$. In terms of classifying spaces, a $G$-structure is represented by a lift
\begin{equation*}
\alxydim{@C=4em}{ & \mathrm{B}G \ar[d]^{\mathrm{B}\rho} \\ M \ar@/^0.8pc/@{-->}[ur] \ar[r]_-{\xi_{\gl M}} & \mathrm{B}\gl d}
\end{equation*}   
of the classifying map of $\mathrm{GL}(M)$ along the induced map $\mathrm{B}\rho$ between classifying spaces. Fundamental questions about $G$-structures include: Does $M$ admit a $G$-structure, and if so, how many? What is the significance of a $G$-structure for the geometry of the manifold?

An important class of examples arises from the homomorphisms
\begin{equation}
\label{spin-sequence}
\spin d \to \so d \to \o d \to \gl d.
\end{equation}
Their geometric meanings, respectively from right to left, are Riemannian metric, orientation, and spin structure. The groups in \cref{spin-sequence} are not arbitrarily chosen; they follow a pattern: Each step in the sequence \quot{kills} the lowest non-vanishing homotopy group of the preceding step (at least in the \quot{stable range} when $d$ is sufficiently high). For instance, $\spin d$ is simply-connected and thus kills the homotopy group $\pi_1(\so d)=\mathbb{Z}_2$, while all higher homotopy groups of $\spin d$ and $\so d$ coincide.

String structures correspond to the next element in the sequence \cref{spin-sequence}, namely the \emph{string group}
\begin{equation*}
\str d \to \spin d,
\end{equation*}
killing the homotopy group $\pi_3(\spin d)=\mathbb{Z}$. In other words, the string group is the 3-connected covering group of the spin group. Consequently, a string structure on $M$ is represented by a principal $\str d$-bundle with an equivariant bundle map to the frame bundle of $M$. 

It is insightful to discuss the obstruction against string structures as a tower of consecutive obstructions:
\begin{equation*}
\alxydim{@R=3em}{
  & \mathrm{B}\str d  \ar[d] \\
  & \mathrm{B}\spin d  \ar[d] \ar[r]^-{\frac{1}{2}\mathrm{p}_1} & \mathrm{B}^4\mathbb{Z} \\
  & \mathrm{B}\so d \ar[r]^-{\mathrm{w}_2} \ar[d] & \mathrm{B}^2\mathbb{Z}_2 \\
  M \ar@{-->}@/^1pc/[ur]\ar@{-->}@/^1pc/[uur]\ar@{-->}@/^1pc/[uuur] \ar[r]_-{\xi_{\o M}} & \mathrm{B}\o d \ar[r]^-{\mathrm{w}_1} & \mathrm{B}\mathbb{Z}_2
}
\end{equation*} 
The right-hand side projections represent the connecting maps of the Puppe sequences associated to the defining fiber sequences of the groups (or more precisely, their classifying spaces). Thus, the classifying map $\xi_{\o M}$ lifts to the next stage of the tower if its composition with the connecting map is null-homotopic.

For example, the defining fiber sequence for the spin group is
\begin{equation*}
\mathrm{B}\Z_2 \to \mathrm{B}\spin d \to \mathrm{B}\so d\text{,}
\end{equation*}
and its connecting map is 
\begin{equation*}
\mathrm{w}_2\in [\mathrm{B}\so d,\mathrm{B}^2\mathbb{Z}_2]=\mathrm{H}^2(\mathrm{B}\so d,\mathbb{Z}_2),
\end{equation*}
known as the \emph{second Stiefel-Whitney class}. Thus, a manifold $M$ with orientation $\xi:M \to \mathrm{B}\so d$ can be equipped with a spin structure if and only if $\mathrm{w}_2 \circ \xi$ is null-homotopic, i.e., if $\mathrm{w}_2(M):=\xi^{*}\mathrm{w}_2\in \mathrm{H}^2(M,\mathbb{Z}_2)$ vanishes. 

Similarly, the definition of the string group is to sit in the fiber sequence
\begin{equation*}
\mathrm{B}^3\Z \to \mathrm{B}\str d  \to \mathrm{B}\spin d\text{,}
\end{equation*}
whose connecting map is the \emph{first fractional Pontryagin class}
\begin{equation*}
\textstyle\frac{1}{2}\mathrm{p}_1 \in [\mathrm{B}\spin d,\mathrm{B}^4\mathbb{Z}] = \mathrm{H}^4(\mathrm{B}\spin d,\mathbb{Z}),
\end{equation*}
a characteristic class of the spin group, denoted by $\textstyle\frac{1}{2}\mathrm{p}_1$ because its double, $2\cdot \textstyle\frac{1}{2}\mathrm{p}_1$, is the pullback of the first Pontryagin class $\mathrm{p}_1\in \mathrm{H}^4(\mathrm{B}\so d,\mathbb{Z})$. It is crucial to note that $\textstyle\frac{1}{2}\mathrm{p}_1$ itself is \emph{not} a pullback of any class in $\mathrm{B}\so d$, and also that the vanishing of $\mathrm{p}_1(M)$ does not imply the vanishing of $\textstyle\frac{1}{2}\mathrm{p}_1(M)$. 

In summary, a \emph{string structure} on a smooth manifold $M$ is a lift of the classifying map of its frame bundle all the way to $\mathrm{B}\str d$, and string structures exist if and only if
\begin{equation*}
\mathrm{w}_1(M)=0, \quad \mathrm{w}_2(M)=0, \quad \textstyle\frac{1}{2}\mathrm{p}_1(M)=0.
\end{equation*} 
Notably, every 3-dimensional spin manifold admits a string structure.
Other examples of string manifolds are described by \cite{Douglas} and \cite{Chen2006}.  

A \emph{string manifold} is a smooth manifold equipped with a string structure. String manifolds form the bordism groups $\Omega^{\mathrm{String}}_{n}$. We remark that in degrees $n \geq 3$, these  coincide with the homotopy groups $\pi_n(\mathrm{M}\mathrm{String})$ of the Thom spectrum $\mathrm{M}\mathrm{String}$ based on stable normal bundles (in contrast to our earlier discussion involving unstable tangent bundles). The first six bordism groups $\Omega^{\mathrm{String}}_{n}$ coincide with the well-known framed bordism groups. The groups between 7 and 16 have been computed by \cite{Giambalvo1971}. Subsequent computations have been carried out by \cite{Hovey1995} and \cite{Mahowald1995}.

There are numerous motivations for introducing tangential structures on manifolds. A compelling motivation for spin structures is to describing (uncharged, massless) fermionic particles in a spacetime $M$. The wave function of these particles is modeled as a section in a vector bundle associated to the frame bundle through a specific representation. This representation is crafted at the Lie algebra level in accordance with physical requirements,  the \quot{spin-statistics theorem}.
This representation does not integrate to a representation of the Lie group $\so d$ but only to one of its covering group, $\spin d$. Consequently, the structure group of $M$ necessitates a lift: a spin structure.

The world-line formalism of the fermionic particle brings about the same requirement, albeit through a different avenue. In this framework, the Feynman path integral sums an exponentiated action functional over all trajectories $\gamma:S^1 \to M$ of a particle. The integral comprises both a bosonic and a fermionic part, with the integrand of the latter being expressed as
\begin{equation*}
\mathcal{A}^{\mathrm{ferm}}_{\gamma}(\psi):=\exp \left ( \int_{S^1} \left \langle \psi,\slashed D_{\gamma}\psi  \right \rangle   \right ),
\end{equation*}
where $\slashed D_{\gamma}$ is the Dirac operator on the world-line $S^1$, twisted by the pullback $\gamma^{*}TM$ of the tangent bundle. The fermionic path integral is the (a priori ill-defined) integral of $\mathcal{A}^{\mathrm{ferm}}_{\gamma}(\psi)$ over all elements $\psi$ in the infinite-dimensional Hilbert space $H := L^2(S^1) \otimes \gamma^{*}TM$; however, it can be rigorously interpreted as a Berezinian integral. In this interpretation, it obtains a well-defined meaning as an element in a one-dimensional complex vector space $\Pfaff(\slashed D)_{\gamma}$. These vector spaces, dependent on the trajectory $\gamma$, collectively form a complex line bundle $\Pfaff(\slashed D)$ over the loop space $LM:=C^{\infty}(S^1,M)$, referred to as the \emph{Pfaffian line bundle} of the family of Dirac operators $\gamma \mapsto \slashed D_{\gamma}$. The phenomenon where the integrand of a path integral is not a function but a section is known as an \emph{anomaly} \citep{freed5}. To cancel this anomaly, a trivialization of the Pfaffian line bundle is required, and it turns out that such a trivialization is furnished by a spin structure.

The narrative outlined above generalizes to string theory, revealing that an anomaly-free superstring necessitates a string structure on $M$. A detailed discussion on this topic is provided in \cref{bunkes-theorem}. Surprisingly, it emerges that a string structure alone is not sufficient to cancel the anomaly. Instead, it must be extended to a \emph{geometric string structure}, which   incorporate genuine geometric information.

In summary, geometric string structures play a pivotal role in the cancellation of anomalies in superstring theory. String structures, and  various versions thereof, have further applications in Mathematical Physics, particularly in M-Theory, as explored in \cite{Sati2012,Sati2019,Saemann2020,Fiorenza2020}.

Moreover, string structures have garnered independent mathematical interest, especially in the context of studying the Witten genus of manifolds. This interest extends to related topics such as the Dirac operator on loop space and the Stolz conjecture. A more in-depth exploration of these mathematical aspects will be presented in \cref{Witten-genus}.

A crucial prerequisite for employing string structures, especially in contexts like anomaly cancellation, involves understanding these structures from a more geometric perspective than through homotopy classes of maps. This becomes particularly essential when delving into the realm of \emph{geometric} string structures. Addressing this issue constitutes a significant aspect of this article, and multiple sections are dedicated to it.

The primary challenge lies in the topological group $\str d$, which exhibits rather unruly behavior. It is not a finite-dimensional Lie group and lacks \quot{good} models. Consequently, the obvious approach  of using principal $\str d$-bundles (along with connections on them) is not suitable.

We explore three alternative yet equivalent definitions of string structures, all of which allow for extensions to geometric string structures:

1.) In \cref{string-structures}, we represent the obstruction class $\frac{1}{2}\mathrm{p}_1(M)$ using \emph{bundle gerbes} and subsequently aim to trivialize them geometrically. Bundle gerbes serve as a well-established tool for representing cohomology classes in higher degrees, akin to how complex line bundles represent cohomology classes in degree two. They also possess a good theory of connections, which we apply in \cref{string-connections} to introduce geometric string structures. Additionally, bundle gerbes align seamlessly with the motivation from string theory, as connections on bundle gerbes model the B-field in string theory. In \cref{bundle-gerbes}, we provide a concise and approachable introduction to bundle gerbes.

2.) In \cref{spin-structures-on-loop-space}, we present a loop space perspective on string structures, where they manifest as \quot{spin structures on loop space}. This viewpoint is strongly motivated by string theory and can be interpreted as the \quot{transgression} of the approach discussed in 1.) to the loop space. Transgression has an elegant implementation utilizing bundle gerbes, as elucidated in \cref{transgression}.

 3.) In \cref{string-2-group}, the string group, which hasn't explicitly featured in the previous two approaches, has a comeback in the guise of a higher-categorical group,  a 2-group. This resurgence is not entirely unexpected, given that string theory is replete with higher-categorical structures. For instance, bundle gerbes may be regarded as categorified line bundles. Extending this notion, categorified bundles for the string 2-group can be employed to comprehend string structures.

In the realm of string theory, the loop space $LM=C^{\infty}(S^1,M)$ comes to the forefront (e.g., as discussed in point 2.) as it serves as the configuration space for closed strings: a point in $LM$ is a string in $M$. This perspective suggests that string theory can be viewed as a theory for point-like particles in the spacetime $LM$. This viewpoint gains traction as the theory of point-like particles is well-understood, as exemplified by the work of \cite{Witten1982}, \cite{atiyah2}, and \cite{killingback1}.

However, the loop space perspective introduces two primary challenges. Firstly, the loop space is infinite-dimensional. Beyond posing numerous analytical problems, it implies that the frame bundle has no $\gl d$-structure for any $d$, consequently disallowing orientations or spin structures in the classical sense. Secondly, the trajectory of a particle in $LM$ inherently possesses the topology of a cylinder in $M$, whereas string theory necessitates to consider  surfaces with arbitrary topology. This challenge can be addressed through the utilization of a technique known as loop-fusion (see \cref{transgression}).

In practice, employing all three perspectives described above simultaneously has proven to be instructive and successful. The amalgamation of higher-categorical geometry and infinite-dimensional analysis has given rise to the term \quot{String geometry}. This contribution aims to provide an introduction and overview of the state of the art in this  field.

\section{Bundle gerbes}

\label{bundle-gerbes}

Cohomology classes in $\h^2(M,\Z)$ can be geometrically represented through   (complex) line bundles. The advantage of such a representation lies in the ability to consider automorphisms of line bundles. Similarly, classes in degree three can be represented by \emph{bundle gerbes}. Given the relevance of this representation to the discussion of (geometric) string structures in \cref{string-structures,string-connections}, as well as their connection to loop spaces, elaborated in \cref{transgression}, we will revisit this representation in the following.

\begin{definition}\citep{murray}
\label{bundle-gerbe}
A \emph{bundle gerbe} $\mathcal{G}$ over a smooth manifold $M$ consists of a surjective submersion $\pi:Y \to M$, a line bundle $L$ over the double fibre product $Y^{[2]} := Y \times_M Y$, and a family of linear maps
\begin{equation*}
\mu_{y_1,y_2,y_3}: L_{y_2,y_3} \otimes L_{y_1,y_2} \to L_{y_1,y_3}\text{,}
\end{equation*} 
parameterized by points $(y_1,y_2,y_3)\in Y^{[3]}$ in the triple fibre product, forming a smooth bundle morphism, and satisfying the evident associativity condition over 4-tuples $(y_1,..,y_4)\in Y^{[4]}$.  
\end{definition}

It is instructive to view the surjective submersion $\pi$ as a generalization of (the disjoint union of) an open cover.
The \emph{trivial bundle gerbe} $\mathcal{I}$ is given by $Y:=M$, $\pi=\id_M$, $L$ the trivial line bundle, and $\mu$ the multiplication of complex numbers. 

Bundle gerbes over $M$ form a symmetric monoidal bicategory $\ugrb M$ with the following properties; see, e.g., \cite{murray,stevenson1,waldorf1,schweigert2}: 
\begin{enumerate}[(i)]

\item
The trivial bundle gerbe $\mathcal{I}$ is the tensor unit. 

\item 
The group of isomorphism classes of objects is isomorphic to $\h^3(M,\Z)$ -- the class corresponding to a bundle gerbe $\mathcal{G}$ is called its \emph{Dixmier-Douady class} and denoted by $\mathrm{dd}(\mathcal{G})$.

\item
The Picard groupoid of automorphisms of any bundle gerbe is  $\ubun M$, the symmetric monoidal category of line bundles over $M$. 

\item
The assignment $M \mapsto \ugrb M$ is a sheaf over the site of smooth manifolds: this means that one can pullback bundle gerbes and glue them together from locally defined ones.

\end{enumerate}
We remark that the definition of bundle gerbes can be motivated by the requirement that   (iv) holds; see  \cite{nikolaus2}.
The surjective submersion appears there as a \quot{covering} in the Grothendieck topology on the category of smooth manifolds. 

If $G$ is a Lie group of Cartan type (compact, connected, simple, simply-connected), then $\h^3(G,\Z)\cong \Z$, and a generator is represented by the \emph{basic gerbe} $\mathcal{G}_{bas}$ over $G$.
It can be constructed explicitly by Lie-theoretic methods \citep{meinrenken1,gawedzki1}.
The basic gerbe $\mathcal{G}_{bas}$ is multiplicative in the sense that it comes with an isomorphism
\begin{equation*}
\mathcal{M}: \pr_1^{*}\mathcal{G}_{bas} \otimes \pr_2^{*}\mathcal{G}_{bas} \to m^{*}\mathcal{G}_{bas}
\end{equation*}
of bundle gerbes over $G \times G$, where $\pr_i$ and $m$ are the projections and multiplication, respectively. Moreover, $\mathcal{M}$ is \quot{coherently associative}.
In general, a multiplicative structure lifts the Dixmier-Douady class of a bundle gerbe over a Lie group  along the homomorphism
\begin{equation}
\label{transgression-classifying}
\h^4(\mathrm{B}G,\Z) \to \h^3(G,\Z)\text{,}
\end{equation}  
see \citet{carey4}.
For a Lie group of Cartan type, both groups in \cref{transgression-classifying} are isomorphic to $\Z$, and above map is a bijection, meaning that the multiplicative structure on $\mathcal{G}_{bas}$ exists and is unique. When $G=\spin d$, $d=3$ or $d > 4$, then the class $\frac{1}{2}\mathrm{p}_1$ generates $\h^4(\mathrm{B}G,\Z)$ and hence induces the Dixmier-Douady class of the basic gerbe $\mathcal{G}_{bas}$ over $\spin d$. In other words, $\mathcal{G}_{bas}$ -- as a multiplicative bundle gerbe -- represents the first fractional Pontryagin class $\frac{1}{2}\mathrm{p}_1$. This will be central for our definition of string structures in \cref{string-structures}.

Another class of examples of bundle gerbes arises from lifting problems; this will be relevant for the loop space theory described in \cref{spin-structures-on-loop-space}. 
If $P$ is a principal $G$-bundle over $M$ and 
\begin{equation}
\label{general-central-extension}
1 \to \ueins \to \tilde G \to G \to 1
\end{equation}
is a central extension, then the \emph{lifting gerbe} $\mathcal{L}_P$ is a bundle gerbe  over $M$  \citep{murray} with surjective submersion $P \to M$ and the associated line bundle $L := (\delta^{*}\tilde G) \times_{\ueins} \C$, where $\tilde G$ is regarded as a principal $\ueins$-bundle over $G$, and  
\begin{equation}
\label{delta-map}
\delta: P^{[2]} \to G,\qquad p\delta(p,p')=p'\text{.}
\end{equation}
The line bundle morphism $\mu$ is induced from the group structure of $\tilde G$. 
The lifting gerbe $\mathcal{L}_P$ comes with an equivalence of categories
\begin{equation}
\label{lifting-theorem}
\Iso_{\ugrb M}(\mathcal{L}_P,\mathcal{I}) \cong \tilde G\text{-}\mathscr{L}\mathrm{ifts}(P)\text{.}
\end{equation}
between the trivializations of $\mathcal{L}_P$, i.e., the isomorphisms between $\mathcal{L}_P$ and the trivial bundle gerbe $\mathcal{I}$, and the category of principal $\tilde G$-bundles $\tilde P$ that lift the structure group of $P$, as explained in \cref{introduction}. 
Hence, the lifting gerbe not only represents the obstruction against lifts (in the sense that it is trivializable if and only if lifts exist), but its trivializations themselves can be identified with the lifts. An illustrative example of this scenario is found in the case of spin$^c$ structures on $M$: consider $P=\so M$ as the frame bundle of an oriented Riemannian manifold $M$ and $\tilde G=\spinc d$; then, spin$^{c}$ structures on $M$ correspond precisely to the trivializations of the lifting gerbe $\mathcal{L}_{\so M}$. The  full theory of bundle gerbes (in which  $L$ may have other structure groups) also extends to the treatment of spin structures or arbitrary tangential structures in a similar fashion.

\section{String structures}

\label{string-structures}

We provide a geometric definition of a string structure employing bundle gerbes, a formulation well-suited for exploring the loop space perspective in \cref{transgression} and their extension to geometric string structures in \cref{string-connections}.

In this section and throughout the subsequent sections,  we assume that $M$ is  a spin manifold of dimension $d$, and we consider the spin structure as a principal $\spin d$-bundle $\spin M$ over $M$. 
We recall that the basic gerbe $\mathcal{G}_{bas}$ over $\spin d$ represents -- as a multiplicative bundle gerbe -- the universal characteristic class $\frac{1}{2}\mathrm{p}_1\in \h^4(\mathrm{B}\spin d,\Z)$. As $\frac{1}{2}\mathrm{p}_1(M)$ is the pullback of $\frac{1}{2}\mathrm{p}_1$ along the classifying map $M \to \mathrm{B}\spin d$ of  $\spin M$, our task is to geometrically implement this pullback. We now elucidate this procedure using bundle 2-gerbes; in \cref{iso-classes-of-string-structures,string-classes}, we present simplifications that do not rely on bundle 2-gerbes.

\begin{definition}\citep{stevenson2}
\label{bundle-2-gerbe}
A \emph{bundle 2-gerbe} $\mathscr{C}$ over $M$ consists of a surjective submersion $\pi:Y \to M$, a bundle gerbe $\mathcal{H}$ over $Y^{[2]}$, a bundle gerbe isomorphism $\mathcal{M}:\pr_{23}^{*}\mathcal{H} \otimes \pr_{12}^{*}\mathcal{H} \to \pr_{13}^{*}\mathcal{H}$ over $Y^{[3]}$ that is \quot{coherently associative}.
\end{definition}

We note that we are employing a slight abbreviation here, overlooking the distinction that being coherently associative is a structure rather than a property.

If $P$ is a principal $G$-bundle over $M$, and $\mathcal{G}$ is a multiplicative bundle gerbe over $G$,
then we may form the \emph{Chern-Simons 2-gerbe} $\mathscr{CS}_P(\mathcal{G})$ \citep{carey4} whose surjective submersion is the bundle projection $P \to M$ and whose bundle gerbe is $\mathcal{H} := \delta^{*}\mathcal{G}$, where $\delta:P^{[2]} \to G$ is the map \cref{delta-map}. The bundle gerbe isomorphism $\mathcal{M}$ can be induced from the multiplicative structure of $\mathcal{G}$. The similarity between this construction and the definition of the lifting gerbe is not coincidental: the Chern-Simons 2-gerbe is a lifting gerbe in a higher sense, see \cref{string-2-group}.

\cite{stevenson2} proves that bundle 2-gerbes have a 2-Dixmier-Douady class in $\h^4(M,\Z)$ and that they are classified up to isomorphism by this class. An important fact, which implicitly appears in \cite{carey4} is the following.

\begin{lemma}
The 2-Dixmier-Douady class of the Chern-Simons 2-gerbe $\mathscr{CS}_P(\mathcal{G})$ is the pullback of the  class of $\mathcal{G}$ in $\h^4(\mathrm{B}G,\Z)$ along the classifying map of $P$. \end{lemma}

In particular, if $M$ is a spin manifold, and $\mathcal{G}_{bas}$ is the basic gerbe over $\spin d$,
then\begin{equation*}
\mathscr{CS}(M):=\mathscr{CS}_{\spin M}(\mathcal{G}_{bas})
\end{equation*}
represents the obstruction class $\frac{1}{2}\mathrm{p}_1(M)$ against string structures on $M$.

Next we discuss trivializations of the Chern-Simons 2-gerbe $\mathscr{CS}(M)$ in order to parameterize the vanishing of this obstruction class.  

\begin{definition}
\citep{stevenson2}
 A \emph{trivialization} of a bundle 2-gerbe $\mathscr{C}=(Y,\mathcal{H},\mathcal{M})$ is a triple $\mathscr{T}=(\mathcal{S},\mathcal{A},\sigma)$ consisting of a bundle gerbe $\mathcal{S}$ over $Y$,  a bundle gerbe isomorphism 
\begin{equation*}
\mathcal{A}: \mathcal{H} \otimes \pr_1^{*}\mathcal{S} \to \pr_2^{*}\mathcal{S}
\end{equation*}
over $Y^{[2]}$, and a 2-isomorphism 
\begin{equation*}
\alxydim{@C=3.5em}{\pr_{23}^{*}\mathcal{H} \otimes \pr_{12}^{*}\mathcal{H} \otimes \pr_1^{*}\mathcal{S} \ar[d]_{\mathcal{M} \otimes \id} \ar[r]^-{\id \otimes \mathcal{A}} & \pr_{23}^{*}\mathcal{H} \otimes \pr_2^{*}\mathcal{S} \ar@{=>}@/_0.5pc/[dl]^*+{\sigma} \ar[d]^{\mathcal{A}} \\ \pr_{13}^{*}\mathcal{H} \otimes \pr_{1}^{*}\mathcal{S} \ar[r]_{\mathcal{A}} & \pr_3^{*}\mathcal{S}}
\end{equation*}
over $Y^{[3]}$ that is compatible with the coherent associativity of $\mathcal{M}$.     
\end{definition}   

\cite{stevenson2} established that a bundle 2-gerbe admits a trivialization if and only if its 2-Dixmier-Douady class vanishes. This result serves as motivation and justification for the ensuing definition, which will form the foundation of our discussion in this article.

\begin{definition}
\citep{waldorf8}
\label{string-structure}
A \emph{string structure} on a spin manifold $M$ is a trivialization $\mathscr{T}$ of the Chern-Simons 2-gerbe $\mathscr{CS}(M)$. 
\end{definition}

Three facts follow now from the general theory of trivializations of bundle 2-gerbes:
\begin{enumerate}[(a)]

\item
\label{properties-string-structure:a} 
String structures exist iff $\frac{1}{2}\mathrm{p}_1(M)=0$.

\item
\label{properties-string-structure:b} 
 String structures form a bicategory. 

\item
\label{properties-string-structure:c} 
If $\mathscr{T}=(\mathcal{S},\mathcal{A},\sigma)$ is a string structure and $\mathcal{K}$ is a bundle gerbe over $M$, then 
\begin{equation}
\label{action-of-gerbes-on-string-structures}
\mathscr{T} \otimes \mathcal{K} := (\mathcal{S} \otimes p^{*}\mathcal{K},\mathcal{A}\otimes \id,\sigma \otimes \id)
\end{equation}
is a another string structure; here, $p$ is the bundle  projection $\spin M \to M$. 

\item
\label{properties-string-structure:d} 
Any section $s$ into the frame bundle $\spin M$ induces a string structure $\mathscr{T}_s$ whose bundle gerbe is $\mathcal{S}=\delta_s^{*}\mathcal{G}_{bas}$, where $s(p(q))=q\cdot \delta_s(q)$. 

\end{enumerate}
We may view \cref{properties-string-structure:c} as an \quot{action} of bundle gerbes on string structures. Going to equivalence classes of string structures -- using \cref{properties-string-structure:b} -- this action has the following  property, see \cite{waldorf8}.  
\begin{lemma}
\label{free-and-transitive-action}
The set of equivalence classes of string structures on a spin manifold $M$ carries a free and transitive action of $\h^3(M,\Z)$. 
\end{lemma}

If one is not interested in the bicategorical structure, one may pass to the homotopy category  (identifying 2-isomorphic 1-morphisms).  In doing so, one can  omit the 2-isomorphism $\sigma$ from  a string structure $\mathscr{T}=(\mathcal{S},\mathcal{A},\sigma)$, and simply retain the knowledge of its existence. If one is not even interested in the 1-categorical structure, a further transition to the set of equivalence classes of objects is feasible. In this case, even the isomorphism $\mathcal{A}$ can be forgotten, as long as  its existence is kept \citep{Waldorfb}. We summarize this in the following.

\begin{lemma}
\label{iso-classes-of-string-structures}
The map $\mathscr{T}=(\mathcal{S},\mathcal{A},\sigma) \mapsto \mathcal{S}$ induces a bijection between the set of equivalence classes of string structures on $M$ and the set of isomorphism classes of bundle gerbes $\mathcal{S}$ over $\spin M$ admitting an isomorphism 
\begin{equation*}
\delta^{*}\mathcal{G}_{bas} \otimes \pr_1^{*}\mathcal{S} \cong \pr_2^{*}\mathcal{S}\text{.}
\end{equation*}
\end{lemma}

We may also look at this picture in a purely cohomological way, by employing the Dixmier-Douady class: the bundle gerbe $\mathcal{S}$ becomes a class $\xi\in \h^3(\spin M,\Z)$, and the existence of the isomorphism $\mathcal{A}$ becomes  the equation 
\begin{equation}
\label{equation-for-string-classes}
\delta^{*}\gamma + \pr_1^{*}\xi = \pr_2^{*}\xi\text{,}
\end{equation}
where $\gamma$ is the Dixmier-Douady class of the basic gerbe $\mathcal{G}_{bas}$, a generator of $\h^3(\spin d,\Z)=\Z$. \Cref{equation-for-string-classes} in fact equivalent to the condition that the restriction of $\xi$ to any fibre of $\spin M \to M$ is a generator (the fibres are diffeomorphic to $\spin d$).
Classes $\xi\in \h^3(\spin M,\Z)$ satisfying this condition are called \emph{string classes}, these have been used by  \cite{stolz4}, \cite{stolz1}, and \cite{redden1} as an alternative notion of string structures.   We have the following result.

\begin{theorem}
\label{string-classes}
\begin{enumerate}[(a)]
\item 
The map 
$\mathscr{T}=(\mathcal{S},\mathcal{A},\sigma) \mapsto \mathrm{dd}(\mathcal{S})$
induces a bijection between equivalence classes of string structures and string classes {\normalfont\citep{waldorf8}.} 

\item
The set of string classes is in bijection with equivalence classes of tangential $\str d$-structures as explained in \cref{introduction} {\normalfont\citep{redden1}}.

\end{enumerate}
\end{theorem}

\Cref{string-classes} expresses the compatibility between string structures in the sense of \cref{string-structure} and tangential string structures.  An alternative approach to establish this compatibility is through the framework of categorified groups, as elucidated in \cref{string-2-group}.

As an example, we consider  the 3-sphere $S^3$, which we identify with the Lie group $\su 2$. Its frame bundle has a canonical, left-invariant trivialization, and hence also a canonical spin structure $\spin {S^3}$ with a canonical section $s$. It induces a  string structure $\mathscr{T}_{can}$ according to \cref{properties-string-structure:d} above. 
If $\mathcal{K}$ denotes the basic gerbe over  $\su 2$,
 we may  consider $\mathscr{T}_{bas}:=\mathscr{T}_{can}\otimes \mathcal{K}$ as another string structure on $S^3$. \cite{bunke2} show that $(S^3,\mathscr{T}_{bas})$ generates the bordism group $\Omega_3^{\mathrm{String}} \cong \Z_{24}$. 

\section{Space/Loop space duality}

\label{transgression}

Essential for the loop space perspective on string structures is a duality between bundle gerbes on $M$ and line bundles on the loop space $LM:=C^{\infty}(S^1,M)$. This duality is established through transgression/regression operations. In this section, we provide an overview of these techniques.

The line bundles on $LM$ involved in the aforementioned duality come equipped with an additional structure,  a \emph{fusion product}. To introduce this concept, we delve into some details about path spaces.
We denote by $PM$ the space of paths $\gamma: [0,1] \to M$. Similar to $LM$, $PM$ can be considered either as a Fr\'echet manifold (with the usual requirement that paths are \emph{flat}, i.e., their derivatives vanish in all orders at the end points) or as a diffeological space (with the usual requirement that paths have \emph{sitting instants}, i.e., they are locally constant near the end points). Both variations allow  the concatenation of paths whenever one ends where the next begins, without loosing smoothness.

We consider the evaluation-at-the-end-points map $PM \to M \times M$ and its fibre products $PM^{[k]}$, which consist of $k$-tuples of paths with common end points. We consider the smooth map
\begin{equation}
\label{cup}
PM^{[2]} \to LM\text{,} \quad (\gamma_1, \gamma_2) \mapsto \gamma_1\cup\gamma_2\text{,}
\end{equation}
producing a loop that first passes along $\gamma_1$ and then retraces backward along $\gamma_2$. Under this map, we view loops as decomposed into two halves.

\begin{definition}
\citep{stolz3}
Let $L$ be a line bundle over $LM$. A \emph{fusion product} on $L$ is a family of linear maps
\begin{equation*}
\lambda_{\gamma_1,\gamma_2,\gamma_3}: L_{\gamma_2 \cup \gamma_3} \otimes L_{\gamma_1 \cup \gamma_2} \to L_{\gamma_1\cup \gamma_3}
\end{equation*}
parameterized by elements  $(\gamma_1,\gamma_2,\gamma_3)\in PM^{[3]}$, forming a smooth bundle morphism and satisfying the evident associativity condition for four  paths. 
\end{definition}  

A point in $PM^{[3]}$ can be interpreted as a \quot{thin pair of pants} -- a pair of paths without a surface, with incoming loops $\gamma_1\cup\gamma_2$ and $\gamma_2\cup\gamma_3$ and an outgoing loop $\gamma_1\cup\gamma_3$:
\begin{center}
\includegraphics[scale=0.8]{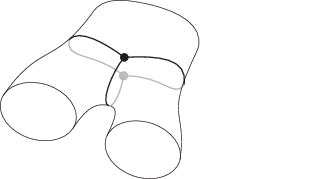}
\setlength{\unitlength}{1pt}\begin{picture}(-30,0)(185,63)\put(73.13357,122.18170){$\gamma_1$}\put(130.45017,97.67282){$\gamma_3$}\put(85.72195,96.65943){$\gamma_2$}\end{picture}
\end{center}
When $\phi:\Sigma \to M$ is a compact oriented surface, such as the worldsheet of a string in $M$, it can be represented from the loop space perspective through a pair-of-pants decomposition, involving combinations of paths in $LM$ (cylinders in $M$) and points in $PM^{[3]}$ (thin pairs of pants). In this sense, fusion products compensate for the limitation of line bundles on the loop space to only represent cylinder-shaped string worldsheets. A comprehensive discussion of this topic can be found in \cite{waldorf10}.

The reader may have observed the resemblance between a fusion product $\lambda$ and the product $\mu$ in the definition of a bundle gerbe (\cref{bundle-gerbe}). This similarity is leveraged by the  \emph{regression} functor, which we now review.

Let $L$ be a line bundle over $LM$ with fusion product $\lambda$. Fixing a point $x\in M$, we define a bundle gerbe $\mathcal{R}_x(L,\lambda)$ with  surjective submersion $\ev_1: P_xM \to M$, the end-point evaluation restricted to the subset of paths starting at $x$. The line bundle is the pullback of $L$ along \cref{cup}, and $\mu:=\lambda$. Denoting by $\ufusbun{LM}$ the category of line bundles over $LM$ with fusion products and fusion-preserving bundle morphisms, we obtain a functor
\begin{equation}
\label{regression}
\mathcal{R}_x: \ufusbun{LM} \to \hc 1\ugrb M\text{,}
\end{equation}
where $\hc 1 \ugrb M$ denotes the homotopy category.

The regression functor \cref{regression} can be turned into an equivalence of categories, by employing  
two modifications of the category $\ufusbun{LM}$: adding equivariance w.r.t. \emph{thin homotopies} of loops, and identifying homotopic bundle morphisms \citep{waldorf11}.

The inverse of regression is called \emph{transgression}. Transgression requires \emph{connections} on bundle gerbes. Historically, it was discovered before regression and discussed in slightly different contexts by \citet{gawedzki3} and \citet{brylinski1}. As connections on bundle gerbes will be required later in \cref{string-connections}, we introduce them in some detail.

\begin{definition}\citep{murray}
A \emph{connection} on a bundle gerbe $\mathcal{G}=(Y,P,\mu)$ is a 2-form $B\in \Omega^2(Y)$ and a connection $\omega$ on $P$, such that $\mu$ is connection-preserving and $R^\omega = \pr_2^{*}B-\pr_1^{*}B$. 
\end{definition}

Here,  $R^{\omega}\in \Omega^2(Y^{[2]})$ denotes the curvature 2-form of the connection $\omega$. 

Every connection on a bundle gerbe $\mathcal{G}$ induces a unique 3-form $R^{\mathcal{G}}\in \Omega^3(M)$ -- called the \emph{curvature} -- such that $\pi^{*}R^{\mathcal{G}}=\mathrm{d}B$. The curvature is a de Rham representative of the Dixmier-Douady class. Every bundle gerbe admits a connection, and the set of connections on a bundle gerbe is an affine space \citep{murray,waldorf8}. Connections on the trivial bundle gerbe $\mathcal{I}$ are precisely the 2-forms $B$ on $M$; their curvature is $\mathrm{d}B$. 

The basic gerbe $\mathcal{G}_{bas}$ over $G$ carries a canonical connection whose curvature is the Cartan 3-form $R^{\mathcal{G}_{bas}}= \frac{1}{6}\left \langle \theta \wedge [\theta \wedge \theta]  \right \rangle$, where $\theta$ is the Maurer-Cartan form \citep{meinrenken1,gawedzki1}. The lifting  gerbe $\mathcal{L}_P$ carries a connection   induced by a connection on the bundle $P$ and by a splitting of the Lie algebra sequence associated to the central extension \cref{general-central-extension}; it controls the lifting of connections \citep{gomi3}.

Connections on bundle gerbes have been studied from the beginning because they model the B-field in string theory \cite{gawedzki3}, just like connections on line bundles model the electromagnetic \quot{A-}field for point-like particles. 
In the context of transgression, they may be regarded as auxiliary structure; later, in \cref{string-connections} they assume an independent role.

Bundle gerbes with connection over $M$ form a symmetric monoidal bicategory $\ugrbcon M$ with the following properties \citep{murray,stevenson1,waldorf1}: 
\begin{enumerate}[(i)]

\item
\label{properties-gerbes-with-connections:a}
The trivial bundle $\mathcal{I}_0$ gerbe with the zero connection is the tensor unit. 

\item 
\label{properties-gerbes-with-connections:b}
The group of isomorphism classes of objects is isomorphic to the differential (or Deligne) cohomology group $\hat \h^3(M,\Z)$.

\item
\label{properties-gerbes-with-connections:c}
The Picard groupoid of automorphisms of any bundle gerbe is the symmetric monoidal groupoid $\ubunconflat M$ of line bundles with \emph{flat} connections. 

\end{enumerate}

Returning to the discussion of transgression, the inverse of regression, we set up a functor
\begin{equation}
\label{transgression-functor}
L:\hc 1 \ugrbcon M \to \ubuncon {LM}. 
\end{equation}
If $\mathcal{G}$ is a bundle gerbe with connection over $M$, then the fibre of the line bundle $L(\mathcal{G})$ over a loop $\tau:S^1 \to M$ is composed  of pairs $([\mathcal{T}],z)$ consisting of a 2-isomorphism class $[\mathcal{T}]$ of trivializations  $\mathcal{T}:\tau^{*}\mathcal{G} \to \mathcal{I}_0$ in $\ugrbcon{S^1}$ and a complex number $z\in \C$, subject to the relation that $([\mathcal{T}],z \cdot \mathrm{Hol}(L)) \sim ([\mathcal{T} \otimes L], z)$, where $L$ is a line bundle with (automatically flat) connection over $S^1$, $\mathrm{Hol}(L) \in \C^{\times}$ is its holonomy around the identity loop $S^1 \to S^1$, and $\mathcal{T} \otimes L$ denotes the tensor product of $\mathcal{T}$ with the automorphism $\mathcal{I}_0 \stackrel{L}{\to} \mathcal{I}_0$, see \cref{properties-gerbes-with-connections:a,properties-gerbes-with-connections:c} above. 

On the level of cohomology classes (Dixmier-Douady classes of bundle gerbes, and first Chern classes of line bundles, respectively), transgression realizes the homomorphism
\begin{equation*}
\tau_M:\h^3(M,\Z) \to \h^2(LM,\Z),\quad \xi \mapsto \int_{S^1}\ev^{*}\xi\text{,}
\end{equation*} 
where $\ev: S^1 \times LM \to M$ is the evaluation of a loop at a time. 
The same formula holds for the curvatures. Transgression also covers a similar homomorphism in differential cohomology \citep{gomi2,Baer2014}.

The line bundles $L(\mathcal{G})$ in the image of transgression are equipped with fusion products  that are compatible with the connections; taking these into account, the transgression functor \cref{transgression-functor} becomes an equivalence of categories \citep{waldorf10}.

\section{Loop-spin structures}
\label{spin-structures-on-loop-space}

Loop-spin structures are the counterpart of the string structures we introduced in \cref{string-structures}, under the space/loop space duality of \cref{transgression}. They have been introduced by \cite{killingback1} and \cite{witten2} and have proved to be very insightful.

We consider again a spin manifold $M$ with spin structure $\spin M$. Taking free loops, we obtain a principal $L\spin d$-bundle $L\spin M$ over $LM$. We  can interpret it as the frame bundle of $LM$, motivated by the fact that a tangent vector at  $\tau\in LM$ is a  vector field along $\tau$. Seeking for an analogy with spin structures, we look for extensions of its structure group, the loop group $L\spin d$. Most prominently, it has the \emph{basic central extension} \citep{pressley1}
\begin{equation}
\label{basic-central-extension}
1 \to \ueins \to \widetilde{L\spin d} \to L\spin d \to 1\text{.}
\end{equation}  

\begin{definition}
\citep{killingback1}
\label{loop-spin-structure}
Let $M$ be a spin manifold. 
A \emph{loop-spin structure on} $M$ is a lift of the structure group of $L\spin M$ along the basic central extension. 
\end{definition}

Thus, a loop-spin structure on $M$ is a principal $\widetilde{L\spin d}$-bundle $\widetilde{L\spin M}$ over $LM$ with an equivariant bundle map
\begin{equation}
\label{bundle-map-of-loop-spin-structure}
\varphi: \widetilde{L\spin M} \to L\spin M\text{.}
\end{equation}

Killingback's motivation for introducing  loop-spin structures was  the cancellation of the global anomaly of the fermionic string, as mentioned in \cref{introduction}. However, the actual cancellation mechanism was discovered much later, utilizing string structures in the sense of \cref{string-structures}. A comprehensive explanation of this anomaly cancellation is provided in \cref{bunkes-theorem}.

As discussed in \cref{bundle-gerbes}, associated to a lifting problem as in \cref{loop-spin-structure} is the lifting gerbe $\mathcal{L}_{L\spin M}$ over $LM$, which -- in the present situation -- is  called the \emph{spin lifting gerbe} on $LM$ and denoted by $\mathcal{S}_{LM}$. By \cref{lifting-theorem}, its trivializations are the loop-spin structures, and its Dixmier-Douady class $\mathrm{dd}(\mathcal{S}_{LM})\in \h^3(LM,\Z)$ obstructs their existence. The relation between this class and the obstruction class $\frac{1}{2}\mathrm{p}_1(M)$ against string structures was first found rationally by \cite{carey7}; the full statement is the following.

\begin{theorem}
{\normalfont\citep{mclaughlin1}}
\label{mclaughlins-theorem}
If $M$ is a spin manifold, then \begin{equation*}
\tau_M(\textstyle\frac{1}{2}\mathrm{p}_1(M)) = \mathrm{dd}(\mathcal{S}_{LM})\text{.}
\end{equation*}
In particular, string manifolds are loop-spin. Moreover, if  $M$ is 2-connected and $d > 5$, then $M$ is string if and only if it is loop-spin.  
\end{theorem}

Efforts have been made to enhance the conditions for equivalence, e.g. by \cite{Kuribayashi1998}. However, results of \cite{Pilch1988} indicate that there exists an essential difference between string manifolds and loop-spin manifolds.

This parallels the fact that spin manifolds have oriented loop spaces, but loop-oriented manifolds are not necessarily spin. In this context, the orientability of $LM$ refers to the vanishing of $\tau_M(\mathrm{w}_2(M)) \in \h^1(LM,\Z_2)$. This distinction was noted by \cite{atiyah2} and resolved by \cite{stolz3}, who introduced fusion products, as discussed in \cref{transgression}, precisely for this purpose. We now present a similar solution for loop-spin structures.

A key observation  is that the line bundle $L_{bas}$ over $L\spin d$ associated to the basic central extension \cref{basic-central-extension}  is equipped with a fusion product. This is a consequence of the equality
\begin{equation*}
\tau_G(\mathrm{dd}(\mathcal{G}_{bas})) = \mathrm{c_1}(L_{bas})
\end{equation*}   
between the transgression of the Dixmier-Douady class of the basic gerbe and the first Chern class of $L_{bas}$,
see \cite{pressley1}. It expresses the possibility to set up a model for the basic central extension using transgression, such that $L(\mathcal{G}_{bas})=L_{bas}$  \citep{waldorf5}.

We spell out what a loop-spin structure becomes under the equivalence of \cref{lifting-theorem}, i.e., as a a trivialization of the spin lifting gerbe  $\mathcal{S}_{LM}$. Namely, the map \cref{bundle-map-of-loop-spin-structure} is the projection of  a principal $\ueins$-bundle, whose associated line bundle $T$ is equipped with a line bundle isomorphism 
\begin{equation}
\label{isomorphism-of-loop-spin-structure}
\delta^{*}L_{bas} \otimes \pr_2^{*}T \cong \pr_1^{*}T
\end{equation}
over $L\spin M^{[2]}$. Next, we aim to employ the regression functor \cref{regression} to obtain from $T$ a bundle gerbe $\mathcal{S}$ over $\Spin M$ and from \cref{isomorphism-of-loop-spin-structure} an isomorphism $\delta^{*}\mathcal{G}_{bas} \otimes \pr_2^{}\mathcal{S} \cong \pr_1^{}\mathcal{S}$, collectively forming a string structure (using \cref{iso-classes-of-string-structures}). This necessitates the following additional structure.

\begin{definition}
{\normalfont\citep{Waldorfa}}
\label{fusive-spin-structure}
A \emph{fusive loop-spin structure} on $M$ is a loop-spin structure whose line bundle $T$ over $L\spin M$ is equipped with a fusion product, such that the isomorphism \cref{isomorphism-of-loop-spin-structure} is fusion-preserving.
\end{definition}

Here,  $\delta^{*}L_{bas}$ is equipped with the pullback of the fusion product on $L_{bas}$.  

Using that regression is functorial, monoidal, and natural,   it maps fusive spin structures to string structures. Conversely, using appropriately chosen connections, string structures transgress to fusive spin structures.  This provides the following improvement of  \cref{mclaughlins-theorem}.

\begin{theorem}
{\normalfont\citep{Waldorfa}}
A spin manifold $M$ of dimension $d=3$ or $d>5$ is string if and only if it is fusive loop-spin. 
\end{theorem}

Unfortunately, the improvement of \cref{fusive-spin-structure} is still insufficient to achieve a bijection between (isomorphism classes of) string structures on $M$ and (isomorphism classes of) fusive loop-spin structures on $LM$. One possibility is to add a version of thin homotopy equivariance to fusive spin structures, see \cite{Waldorfb}. Another possibility was found by \cite{Kottke} who used a combination of reparameterization equivariance and a certain figure-eight move.

Loop-spin structures are the starting point for the construction of the \emph{spinor bundle on loop space}.
The idea is to use a representation of the basic central extension on a Hilbert space $H$ and  to form the associated Hilbert space bundle
\begin{equation}
\label{spinor-bundle-on-loop-space}
S_{LM} := \widetilde{L\spin M} \times_{\widetilde{L\spin d}} H
\end{equation}
over $LM$. This, and a further involvement  fusion products is discussed in \cref{stringor-bundle}.

\section{The string 2-group}
\label{string-2-group}

The treatment of string structures presented in \cref{string-structures} is motivated and centered around the trivialization of obstruction classes. In the present section, we aim to pursue and geometrically interpret the original approach through tangential structures, elucidated in \cref{introduction}.

As mentioned earlier, the string group $\str d$, defined as a 3-connected cover of $\spin d$, is not a finite-dimensional Lie group. For instance, Stolz' model \citep{stolz4} for $\str d$ takes the form of an extension
\begin{equation*}
F \to \str d \to \spin d
\end{equation*}
of topological groups whose fibre $F$ is a $\mathrm{K}(\Z,2)$. This implies that $F$ has cohomology in infinite degrees. By the Serre spectral sequence,  $\str d$ then also possesses cohomology in infinite degrees, and consequently cannot be modeled by a finite-dimensional manifold. \cite{Nikolausb} later showed that Stolz' model can be endowed with the structure of an infinite-dimensional Fr\'echet Lie group.

More successful, thanks to its profound connections to the physics motivation, was a program by Baez et al. to embed the string group in the context of higher-categorical gauge theory, viewing it as a \emph{2-group}; see, for instance, \cite{baez5,baez2}.

Several models of the \emph{string 2-group} in different contexts have since then been constructed, e.g., as a strict Fr\'echet Lie 2-group by \cite{baez9}, as a simplicial Lie group by \cite{Henriques2008} as a finite-dimensional smooth \quot{stacky} 2-group by \cite{pries2}, as a strict diffeological 2-group \citep{Waldorf}, or as a smooth $\infty$-group by  \cite{Bunk2020}.

We will focus on one of these versions: strict 2-groups. Strict 2-groups are group objects in groupoids or, equivalently, internal groupoids in groups. Specifically, a \emph{strict 2-group} is a groupoid $\Gamma$ where the set of objects $\Gamma_0$ and the set of morphisms $\Gamma_1$ are equipped with group structures, such that the source and target maps $s,t: \Gamma_1\to\Gamma_0$, composition, identity-assigning map $i:\Gamma_0 \to \Gamma_1$, and inversion are all group homomorphisms.

The data of a strict 2-group are somewhat redundant, so it is often convenient to consider a minimal version called a \emph{crossed module}. A crossed module is a group homomorphism $t:H \to K$ together with an action of $K$ on $H$ by group homomorphisms satisfying
\begin{equation}
\label{crossed-module}
t(k\cdot h) = gt(k)g^{-1}
,\quad
t(h)\cdot h'=hh'h^{-1}
\end{equation}
for all $k\in K$ and $h,h'\in H$. When passing from a strict 2-group $\Gamma$ to the corresponding crossed module, we get $G= \Gamma_0$ and $H=\mathrm{ker}(s)\subset \Gamma_1$. Both strict 2-groups and crossed modules have versions with topological groups, (Fr\'echet) Lie groups, or diffeological groups, and they are equivalent in each version; see, for example, \cite{brown1}.

The construction of the string 2-group by \cite{baez9} yields a Fr\'echet crossed module. We describe here a simplification found lately by \cite{Ludewig2023}. We start with a Lie group $G$ and a central extension of its loop group,
\begin{equation*}
1\to \ueins \to \widetilde{LG} \stackrel{p}\to LG \to 1.
\end{equation*} 
To get the string 2-group, we specialize to $G=\spin d$ and the basic central extension \cref{basic-central-extension}.
The group $K$ of the crossed module we want to construct is $K:=P_eG$, the Fr\'echet Lie group of smooth paths $\gamma: [0,\pi] \to G$ that start at the neutral element $e\in G$, and whose derivatives at the endpoints vanish in all orders. The latter condition ensures that one can form the smooth loop $\Delta\gamma \in LG$ which passes along $\gamma$ and then retraces back along same way, i.e., $\Delta\gamma = \gamma \cup \gamma$ in the notation of \cref{transgression}. 
We denote by $L_0G\subset LG$ the subgroup of loops $\tau:S^1 \to G$ with $\tau(t)=e$ for $t\in [\pi,2\pi]$, and let $H$ be the restriction of $\widetilde{LG}$ to that subgroup. The homomorphism $t:H \to K$ is $t(\Phi) := p(\Phi)|_{[0,\pi]}$. The action of $K$ on $H$ is 
\begin{equation}
\label{easy-action}
\gamma \cdot \Phi := \widetilde{\Delta\gamma} \cdot \Phi \cdot \widetilde{\Delta\gamma}^{-1}\text{,}
\end{equation} 
where $\widetilde{\Delta\gamma}\in \widetilde{LG}$ is any lift of $\Delta\gamma\in LG$. It is easy to see that this is well-defined and that the first identity in \cref{crossed-module} is satisfied. 

The second identity in \cref{crossed-module} requires an additional property of the central extension $\widetilde{LG}$, namely, that it is \emph{disjoint-commutative} \cite{Waldorfc}. This means that elements $\Phi_1,\Phi_2\in \widetilde{LG}$ commute whenever their base loops $p(\Phi_1),p(\Phi_2) \in LG$ have disjoint support, i.e., for each $t\in S^1$ either $p(\Phi_1)(t)=e$ or $p(\Phi_2)(t)=e$.
When $G$ is semisimple and simply-connected, every central extension of $LG$ has this property \citep{Ludewig2023}; in particular, the basic central extension \cref{basic-central-extension} is disjoint-commutative. We remark that this is related to nets of operator algebras coming from loop group extensions; see, e.g. \cite{Gabbiani1993}.

Above description, applied to $G=\spin d$ and the basic central extension \cref{basic-central-extension} gives a complete definition of (the crossed module of) the string 2-group, which we will denote by $\mathscr{S}\mathrm{tring} (d)$. It is canonically and strictly isomorphic as Fr\'echet Lie 2-groups to the original construction of \cite{baez9}, where the definition of the action \cref{easy-action} is more complicated.

Strict topological 2-groups induce ordinary topological groups through geometric realization. Geometric realization establishes the relationship between higher gauge theory and ordinary gauge theory, and the subsequent result shows that the construction of the string 2-group, as outlined above, fulfills its intended purpose.

\begin{theorem}
\label{realization-of-string-2-group}
{\normalfont\citep{baez9}}
The geometric realization of the string 2-group $\XString(d)$ is a 3-connected cover of $\spin d$. 
\end{theorem} 

Further developments include  a theory of central extensions of 2-groups \citep{baez9,pries2}. In this framework, the string 2-group becomes a central extension
\begin{equation}
\label{string-extension}
\mathscr{B}\ueins \to \XString(d) \to \Spin d\text{.}
\end{equation}
Here, $\mathscr{B}\ueins$ is the 2-group whose crossed module is $\ueins \to {\ast}$ and whose geometric realization is the classifying space $\mathrm{B}\ueins$. The group $\spin d$ is considered as a (categorically) discrete 2-group.

Higher gauge theory is of course not only concerned with categorified groups but also with categorified principal bundles, known as \emph{principal 2-bundles}, which are bundles for structure 2-groups. Bundle gerbes, introduced in \cref{bundle-gerbes}, are an example of categorified bundles for the structure 2-group $\mathscr{B}\ueins$. Generalizing bundle gerbes to arbitrary structure 2-groups becomes technically involved, so we will not delve into the details here; refer to \cite{Nikolaus} for an overview.
However, we notice that within the theory of principal 2-bundles, one can now discuss lifts of structure 2-groups.

\begin{theorem}
\citep{Nikolaus}
\label{string-structures-and-lifts}
Let $M$ be a spin manifold with spin structure $\Spin M $. Then, a string structure in the sense of \cref{string-structure} is the same as a lift of the structure group of $\Spin M$ along the central  extension \cref{string-extension}.
\end{theorem}   

Under geometric realization, and as a consequence of \cref{realization-of-string-2-group}, lifts of the structure group of $\spin M$ to the string 2-group become $\str d$-principal bundles, i.e., tangential string structures. Therefore, \cref{realization-of-string-2-group} offers an alternative perspective on the equivalence between the string structures of \cref{string-structure} and tangential string structures.

The essence of \cref{string-structures-and-lifts} lies in a  relationship between the string 2-group $\XString(d)$ and the basic gerbe $\mathcal{G}_{bas}$ over $\spin d$. This relationship manifests itself in the fact that the Chern-Simons 2-gerbe $\mathscr{CS}(M)$ serves as a categorified lifting gerbe for the problem of lifting the structure group of $\spin M$ from $\spin d$ to $\XString(d)$. A categorified lifting theorem, akin to \cref{lifting-theorem}, consequently implies that its trivializations are the lifts.

\section{The stringor bundle}

\label{stringor-bundle}

The stringor bundle, by terminology, assumes the role of the spinor bundle but for strings rather than point-like particles. It originates from a proposal by \cite{stolz3} to combine the spinor bundle on loop space $S:=S_{LM}$ from \cref{spin-structures-on-loop-space} with loop-fusion, implementing the ingenious idea that it is the transgression of a (then unknown) geometric structure on $M$ itself.

\cite{stolz3} proposed introducing the following additional structure for the spinor bundle $S$ on loop space:
\begin{enumerate}
\labelsep=0.5em

\item 
A bundle $A$ of von Neumann algebras over the space $PM$ of paths in $M$.

\item
For each loop $\tau = \gamma_1\cup\gamma_2$, the fibre $S_{\tau}$ becomes an $A_{\gamma_2}$-$A_{\gamma_1}$-bimodule. 

\item
For each  $(\gamma_1,\gamma_2,\gamma_3)\in PM^{[3]}$, there is an isomorphism
\begin{equation}
\label{fusion-in-spinor-bundle}
S_{\gamma_2\cup \gamma_3} \boxtimes_{A_{\gamma_2}} S_{\gamma_1\cup \gamma_2} \cong S_{\gamma_1\cup\gamma_3} 
\end{equation} 
of $A_{\gamma_1}$-$A_{\gamma_3}$-bimodules, where $\boxtimes$ denotes the Connes confusion of bimodules. Moreover, the isomorphisms \cref{fusion-in-spinor-bundle} are associative over quadruples of paths.  

\end{enumerate}   
The appearance of von Neumann algebras, and in particular, of the hyperfinite type III$_1$-factor, is attributed  to the spinor bundle $S$ being an infinite-dimensional Hilbert space bundle. Moreover,   there are  relations to another model of the string group employed by \cite{stolz1}, and to the Connes fusion of positive energy representations \citep{Wassermann1998,ToledanoLaredo1997}.

The structure outlined in points 1-3 was rigorously constructed by \cite{Kristel2019,Kristel2020b,Kristel2019b}. The key element was to employ a \emph{fusive} spin structure, as defined in \cref{fusive-spin-structure}.

To provide additional details, the von Neumann algebra bundle $A$ over $PM$ has, as its typical fiber, the hyperfinite type III$_1$-factor $N=\vN$, modeled as the completion of a subalgebra of the Clifford algebra of a Fock space $F$. The Fock space $F$ itself can be identified with a standard form of $N$, an $N$-$N$-bimodule that is neutral with respect to Connes fusion. On the other hand, $F$  carries the representation used to perform the associated bundle construction of $S$ in \cref{spinor-bundle-on-loop-space}. The isomorphism \cref{fusion-in-spinor-bundle} is constructed in such a way that there exist local trivializations $u_{23}$, $u_{12}$, and $u_{13}$ of $S$ in the neighborhood of loops $\gamma_2\cup\gamma_3$, $\gamma_1\cup\gamma_2$, and $\gamma_1\cup\gamma_3$, respectively, such that the diagram
\begin{equation*}
\alxydim{}{S_{\gamma_2\cup \gamma_3} \boxtimes_{A_{\gamma_2}} S_{\gamma_1\cup \gamma_2} \ar[d]_{u_{23} \boxtimes u_{12}} \ar[r] &  S_{\gamma_1\cup\gamma_3} \ar[d]^{u_{13}} \\ F \boxtimes_N F \ar[r] & F}
\end{equation*}   
is commutative.  This diagram involves on the bottom the canonical map expressing the neutrality of  the bimodule $F$ with respect to Connes fusion.

The structure outlined above aligns with a relatively new framework for categorified vector bundles, also known as \emph{2-vector bundles}. In this context, a \emph{2-vector space} is nothing but a (unital, associative) algebra, but  considered as an object in a bicategory, with bimodules serving as morphisms. The corresponding concept of a 2-vector bundle is then derived through stackification, resulting in the following structure akin to a bundle gerbe.

\begin{definition}
\citep{Kristel2020}
\label{2-vector-bundle}
A \emph{2-vector bundle} $\mathscr{V}$ over $M$ consists of a surjective submersion $\pi:Y \to M$, an algebra bundle $A$ over $Y$,  a vector  bundle $M$ over $Y^{[2]}$ in which each fibre $M_{y_1,y_2}$ is an $A_{y_2}$-$A_{y_1}$-bimodule, and a family of intertwiners
\begin{equation*}
\mu_{y_1,y_2,y_3}: M_{y_2,y_3} \otimes_{A_{y_2}} M_{y_1,y_2} \to M_{y_1,y_3}
\end{equation*}
parameterized by points $(y_1,y_2,y_3)\in Y^{[3]}$ , forming a smooth bundle morphism and satisfying the evident associativity condition over 4-tuples $(y_1,..,y_4)\in Y^{[4]}$. 
\end{definition}

Von Neumann algebras are algebras with analytical additional structure, and it is instructive to view them within the realm of 2-vector spaces as \emph{2-Hilbert spaces}. A corresponding theory of 2-Hilbert space \emph{bundles} was developed in \cite{Kristel2022a} and \cite{Ludewig2023a}, where the relative tensor product in \cref{2-vector-bundle} is replaced by  fibre-wise Connes fusion.
This framework provides a solid foundation for the structure outlined above, resembling the regression functor of \cref{transgression}.
\begin{definition}
\label{definition-of-stringor-bundle}
\citep{Kristel2022a}
The \emph{stringor bundle} $\mathscr{S}_M$ is the 2-Hilbert space bundle whose surjective submersion is the end-point evaluation $\ev_1: P_xM \to M$, whose von Neumann algebra bundle is the restriction of the bundle $A$ to $P_xM \subset PM$,  and whose bimodule bundle $M$ is the pullback of the spinor bundle $S$ along the map $\cup: P_xM^{[2]} \to LM$ of \cref{cup}.  
\end{definition}

While \cref{definition-of-stringor-bundle} stands as a valid and rigorous definition, supported by the loop space perspective of string theory, the question arises as to whether the stringor bundle has another presentation that aligns more with the classical definition of the spinor bundle as an associated vector bundle. Indeed, such an alternative presentation exists.

Of central importance is a certain representation of the string 2-group $\XString(d)$. In general, 2-groups have representations on 2-vector spaces, i.e., on algebras. If $N$ is a algebra, then its \emph{automorphism 2-group} $\AUT(N)$ is given by the  crossed module (see \cref{string-2-group}) 
$N^{\times} \stackrel{c}\to \aut(N)$, 
where 
$N^{\times}$ is the group of units of $N$ and $\mathrm{Aut}(N)$ is the group of automorphisms of $N$, $c$ sends a unit to the inner automorphism given by conjugation, and $\aut (N)$ acts on $N^{\times}$ by evaluation. A representation of a 2-group $\Gamma$ on a 2-vector space $N$ is a 2-group homomorphism
$\mathscr{R}:\Gamma \to \AUT(N)$. When $\Gamma$ is described itself by a crossed module $H \stackrel t\to G$, then $\mathscr{R}$ consists of group homomorphisms $\mathscr{R}_0:G \to \mathrm{Aut}(N)$ and $\mathscr{R}_1:H \to N^{\times}$ compatible with the structure maps of the two crossed modules. 

As von Neumann algebras are 2-\emph{Hilbert} spaces, they have a \emph{unitary} automorphism 2-group $\mathscr{U}(N)$, obtained by shrinking the groups of $\AUT(N)$  to the subgroups $\ug N\subset N^{\times}$ of unitaries on $N$      and  $\mathrm{Aut}^{*}(N)$ of $\ast$-automorphisms. 

\begin{definition}
\citep{Kristel2022b}
A \emph{unitary representation} of a strict 2-group $\Gamma$ on a von Neumann algebra $N$ is a 2-group homomorphism
$\mathscr{R}:\Gamma \to \mathscr{U}(N)$. 
\end{definition}

If $\Gamma$ is a \emph{topological} strict 2-group, we typically demand that unitary representations  be  continuous. Therefore, we equip $\ug N$  with the ultraweak topology, and  $\mathrm{Aut}^{*}(N)$  with Haagerup's u-topology.

The string 2-group $\XString(d)$ reviewed in \cref{string-2-group} possesses a canonical, continuous unitary representation on the type III$_1$ factor $N=\vN$, as established in \cite{Kristel2022b}. It consists of group homomorphisms 
\begin{align*}
&\mathscr{R}_0: P_e\spin d \to \aut^{*}(N)
\\
&\mathscr{R}_1: \widetilde{L\spin d}|_{L_0\spin d} \to \ug N
\end{align*}
which are defined are defined utilizing the aforementioned model for $N$ as a Clifford-von Neumann algebra of a Fock space $F$. Essentially, $\mathscr{R}_0$ associates  Bogoliubov automorphisms, while $\mathscr{R}_1$ leverages the fact that $F$ carries a representation of the basic central extension $\widetilde{L\spin d}$.  

Given the representation $\mathcal{R}$ of $\XString(d)$, one can employ a higher analogue of the classical associated bundle construction. It associates to a unitary representation $\mathscr{R}:\Gamma \to \mathscr{U}(N)$ of a topological 2-group $\Gamma$ on a von Neumann algebra $N$ and a principal $\Gamma$-2-bundle $\mathcal{P}$ a 2-Hilbert space bundle denoted as $\mathcal{P} \times_{\Gamma} N$.
We will not provide any more details here and only state the following result. 

\begin{theorem}
{\normalfont\citep{Kristel2022a}}
\label{stringor-bundle-associated}
Let $M$ be a string manifold with a string structure $\mathscr{T}$. Suppose that:
\begin{itemize}

\item 
$\widetilde{L\spin M}$ is the fusive loop-spin structure on $LM$ that corresponds to $\mathscr{T}$ under regression; let $\mathscr{S}_M$ be the corresponding stringor bundle, and  
\item
$\XString(M)$ is a principal $\XString (d)$-2-bundle over $M$ that lifts the structure group of $\spin M$ and corresponds to $\mathscr{T}$  under \cref{string-structures-and-lifts}.

\end{itemize}
Then, there exists a canonical isomorphism of 2-Hilbert space bundles
\begin{equation*}
\mathscr{S}_{M} \cong \XString(M) \times_{\XString(d)} \vN\text{.}
\end{equation*}
\end{theorem}

\Cref{stringor-bundle-associated}  exhibits Stolz-Teichner's stringor bundle $\mathscr{S}_M$ as an associated 2-Hilbert space bundle, thereby establishing the desired analogy between the stringor bundle and  the spinor bundle.  

\section{String connections}
\label{string-connections}

We come back to our description of string structures as trivializations of the Chern-Simons 2-gerbe $\mathscr{CS}(M)$ from \cref{string-structures}, now with the objective of advancing to \emph{geometric} string structures.

\begin{definition}
\label{bundle-2-gerbes-with-connection}
\citep{stevenson2}
A \emph{connection on a bundle 2-gerbe gerbe} $\mathscr{C}=(Y,\mathcal{H},\mathcal{M})$ as in \cref{bundle-2-gerbe} is a 3-form $C \in \Omega^3(Y)$, and a connection on $\mathcal{H}$ such that the isomorphism $\mathcal{M}$ becomes connection-preserving. A \emph{connection on a trivialization} $\mathscr{T}=(\mathcal{S},\mathcal{A},\sigma)$ of $\mathscr{C}$ is a connection on the bundle gerbe $\mathcal{S}$ such that $\mathcal{A}$ and $\sigma$ are connection-preserving.
\end{definition}

We remark that this is again slightly abbreviating: for a bundle gerbe isomorphism (such as $\mathcal{M}$, $\mathcal{A}$), being connection-preserving is structure, not property.

The Chern-Simons 2-gerbe  carries a canonical connection, as established by \cite{carey4}. Its 3-form is the Chern-Simons 3-form
\begin{equation*}
C := \left \langle A \wedge \mathrm{d}A  \right \rangle + \frac{1}{3} \left \langle A \wedge [A \wedge A ] \right \rangle\text{,}
\end{equation*}
hence the name of the bundle 2-gerbe. 
Here, $A$ is the Levi-Civita connection 1-form, and $\left \langle -,-  \right \rangle$ is the basic inner product on the Lie algebra of $\spin d$.
The connection on the bundle gerbe $\mathcal{H}=\delta^{*}\mathcal{G}_{bas}$ of $\mathscr{CS}(M)$ is the pullback connection of the canonical connection on the basic gerbe, shifted by the 2-form 
\begin{equation*}
\omega := \left \langle \delta^{*}\theta\wedge \pr_1^{*}A  \right \rangle\in \Omega^2(\spin M^{[2]})\text{;}
\end{equation*}
in other words, $\mathcal{H}= \delta^{*}\mathcal{G}_{bas} \otimes \mathcal{I}_{\omega}$. The shift is necessary in order to ensure that the isomorphism $\mathcal{M}$ is connection-preserving. 

Every connection on a bundle 2-gerbe $\mathscr{C}$ determines a curvature 4-form $R^{\mathscr{C}}\in \Omega^4(M)$ satisfying the condition $\pi^{*}R^{\mathscr{C}}=\mathrm{d}C$, where $\pi:Y \to M$ is the surjective submersion of $\mathscr{C}$. The 4-form $R^{\mathscr{C}}$ is a de Rham-representative of the 2-Dixmier-Douady class of $\mathscr{C}$. A trivialization $\mathscr{T}=(\mathcal{S},\mathcal{A},\sigma)$ with connection determines a \quot{covariant derivative} 3-form $D^{\mathscr{T}}$ 
satisfying the condition $\pi^{*}D^{\mathscr{T}}=C+R^{\mathcal{S}}$, where $R^{\mathcal{S}}$ is the curvature of the connection on the bundle gerbe $\mathcal{S}$. The covariant derivative satisfies   $\mathrm{d}D^{\mathscr{T}}=R^{\mathscr{C}}$; see \cite{waldorf8}. In case of the Chern-Simons 2-gerbe $\mathscr{CS}(M)$, the curvature 4-form is the Pontryagin 4-form 
\begin{equation*}
R^{\mathscr{CS}(M)} = \frac{1}{2}\left \langle R^{A} \wedge R^{A}  \right \rangle \text{.}
\end{equation*}

\begin{definition}
\citep{waldorf8}
\label{string-connection}
Let $\mathscr{T}$ be a string structure in the sense of \cref{string-structure}. A \emph{string connection} is a connection on $\mathscr{T}$, and the pair of $\mathscr{T}$ and a string connection is called \emph{geometric string structure}.
\end{definition}

According to the general remarks above, each geometric string structure $\mathscr{T}$ determines a 3-form $D^{\mathscr{T}}\in \Omega^3(M)$ such that $\mathrm{d}D^{\mathscr{T}}=\frac{1}{2}\left \langle R^{A} \wedge R^{A}  \right \rangle$. Such a 3-form alone is sometimes referred to as a \emph{rational string structure}. Additional general facts about connections on bundle 2-gerbes imply statements about geometric string structures. For instance, the action of bundle gerbes on string structures (see \cref{properties-string-structure:c} in \cref{string-structures}) lifts to an action $(\mathscr{T},\mathcal{K}) \mapsto \mathscr{T} \otimes \mathcal{K}$ of bundle gerbes \emph{with connection} on \emph{geometric} string structures, given by the same formula, \cref{action-of-gerbes-on-string-structures}.
The covariant derivative satisfies 
\begin{equation}
\label{covariant-derivative-under-action}
D^{\mathscr{T}\otimes \mathcal{K}}=D^{\mathscr{T}}+R^{\mathscr{K}}
\end{equation}
A further fact is the following. 

\begin{theorem}
\label{string-connections-exist}
{\normalfont\citep{waldorf8}}
Every string structure $\mathscr{T}$ admits a string connection, and the set of all string connections on $\mathscr{T}$ forms an affine space. In particular, every string manifold admits a geometric string structure.  
\end{theorem}

Finally, geometric string structures again form a bicategory, whose complexity can be reduced by passing to its homotopy category, or further to its set of isomorphism classes; see \cite{Waldorfb}. For instance,  \Cref{iso-classes-of-string-structures} lifts almost verbatim to geometric string structures:

\begin{lemma}
\label{iso-classes-of-geometric-string-structures}
The map 
$\mathscr{T}=(\mathcal{S},\mathcal{A},\sigma) \mapsto \mathcal{S}$
induces a bijection between the set of equivalence classes of geometric string structures on $M$ and the set of isomorphism classes of bundle gerbes $\mathcal{S}$ with connection over $\spin M$ admitting a connection-preserving isomorphism 
\begin{equation*}
\delta^{*}\mathcal{G}_{bas} \otimes \mathcal{I}_{\omega} \otimes \pr_1^{*}\mathcal{S} \cong \pr_2^{*}\mathcal{S}\text{.}
\end{equation*}
\end{lemma}

At this point, one can also pass to differential cohomology, and consider \emph{differential string classes}, classes $\xi\in \hat \h^3(\spin M)$ such that
\begin{equation}
\label{condition-for-differential-string-classes}
\delta^{*}\hat \gamma  +\hat\omega + \pr_1^{*}\xi = \pr_2^{*}\xi
\end{equation}
holds over $\spin M^{[2]}$, where $\hat\gamma\in \hat\h^3(\spin d)$ is the differential cohomology class of the basic gerbe (with its canonical connection), and $\hat\omega$ is the image of the 2-form $\omega$ under the structure map $\Omega^k(M) \to \hat \h^{k+1}(M)$ of differential cohomology. It is, however, not possible to replace \cref{condition-for-differential-string-classes} by the condition that \quot{$\xi$ restricts to $\hat\gamma$ in each fibre} as we did for string classes, since  in differential cohomology this condition is strictly weaker than \cref{condition-for-differential-string-classes}; see \cite{Waldorfb}.

A remarkable result of \cite{redden1} shows using harmonic analysis that  there is a \emph{canonical} rational string structure $D^{hrm}$ associated to every string structure $\mathscr{T}$, additionally satisfying $\mathrm{d}\ast D^{hrm}=0$. There is, however, no canonical string connection associated to a string structure: the group
of equivalence classes of topologically trivial bundle gerbes with flat connections acts free and transitively on the string connections for fixed $\mathscr{T}$ and fixed covariant derivative $D$; see \cref{covariant-derivative-under-action}.
That group is
\begin{equation*}
\Omega^2_{cl}(M) / \Omega^2_{cl,\Z}(M) \cong \h^2(M,\R)/\h^2(M,\Z)
\end{equation*}
and hence is not automatically trivial.

Our  example of string structures on the 3-sphere from \cref{string-structures} extends to geometric string structures.  We recall from \cref{properties-string-structure:d} in   \cref{string-structures} that any section $s$ into the surjective submersion of a bundle 2-gerbe $\mathscr{C}$ induces a trivialization $\mathscr{T}_s$. If $\mathscr{C}$ carries a connection with 3-form $C$, then $\mathscr{T}_s$ also carries a connection, with covariant derivative $D^{\mathscr{T}_s}=s^{*}C$. Thus, the 3-sphere has a canonical geometric string structure $\mathscr{T}_{can}$. If $\mathcal{K}$ denotes the basic gerbe of the group $\su 2$, equipped with its canonical connection,  then $\mathscr{T}_{bas}:=\mathscr{T}_{can}\otimes \mathcal{K}$ is another geometric string structure on $S^3$.

In the remainder of this section we  describe the loop space perspective to string connections (analogous to \cref{spin-structures-on-loop-space}) and the 2-group perspective (analogous to \cref{string-2-group}).

Concerning the loop space perspective, we recall from \cref{loop-spin-structure} that a loop-spin structure is a principal $\widetilde{L\spin d}$-bundle $\widetilde{L\spin M}$ over $LM$ that lifts the structure group of the looped frame bundle $L\spin M$ along the basic central extension \cref{basic-central-extension}. The Levi-Civita connection $A$ on $\spin M$ can be \quot{looped} to yield a principal connection $\omega_A$ on $L\spin M$. This leads to  the following natural definition.

\begin{definition}
\label{loop-spin-connections}
\citep{Coquereaux1998}
A \emph{loop-spin connection} is a connection $\Omega$ on the principal bundle $\widetilde{L\spin M}$ that lifts the looped Levi-Civita connection $\omega_A$. 
\end{definition}

More precisely, the statement that $\Omega$ lifts $\omega_A$ means that $p_{*}(\Omega) = \varphi^{*}\omega_A$, where $p$ is the projection in the basic central extension, $p_{*}$ is the induced Lie algebra homomorphism, and $\varphi$ is the map \cref{bundle-map-of-loop-spin-structure}.

We recall that  loop-spin structures are related to the string structures of \cref{string-structure} through regression and transgression functors. While in \cref{spin-structures-on-loop-space} the usage of regression was more natural (since it is independent of connections), in the present context, where connections play a crucial role, transgression assumes greater significance. Suppose $\mathscr{T}=(\mathcal{S},\mathcal{A},\sigma)$ is a geometric string structure on $M$ in the sense of \cref{string-connection}. Applying the transgression functor \cref{transgression-functor}, we obtain a line bundle $T := L\mathcal{S}$ over $L\spin M$ and a line bundle isomorphism
\begin{equation*}
L\mathcal{A}: L(\delta^{*}\mathcal{G}_{bas} \otimes \mathcal{I}_{\omega}) \otimes
\pr_1^{*}T \to \pr_2^{*}T
\end{equation*} 
over $L\spin M^{[2]}$, satisfying a coherence condition over $L\spin M$ coming from the existence of $\sigma$. Since the tensor product with $\mathcal{I}_{\omega}$ is only a shift of the connection, the transgression $L(\delta^{*}\mathcal{G}_{bas} \otimes \mathcal{I}_{\omega})$ is, as a bundle, isomorphic to $\delta^{*}L_{bas}$. Thus, $T$ and $L\mathcal{A}$ form a trivialization of the spin lifting gerbe $\mathcal{S}_{LM}$, see \cref{isomorphism-of-loop-spin-structure}, and hence a loop-spin structure.  Incorporating fusion products leads to the ensuing result.  

\begin{theorem}
\label{equivalence-geometric-string-structures-loop-space}
{\normalfont\citep{Waldorfb}}
Transgression establishes an equivalence between geometric string structures and fusive loop-spin structures equipped with fusive, superficial loop-spin connections. 
\end{theorem} 

For brevity, we omit the explanation of the conditions required for loop-spin connections to be fusive and superficial.

It remains to relate the string connections of \cref{string-connection} to connections on categorified principal bundles for the string 2-group. A well-established theory of connections on such bundles was initiated by \cite{breen1} and further justified by the existence of parallel transport along paths and surfaces \citep{schreiber2}. A version for bundle gerbes was developed in \cite{aschieri}, and for principal 2-bundles in \cite{Waldorf2016}. These connections are subject to a condition known as \emph{fake-flatness}.

Unfortunately, fake-flatness poses a significant constraint in the context of string structures, as noted by \cite{sati1}: if $\XString(M)$ is a principal $\XString(d)$-2-bundle lifting  $\spin M$, then imposing that $\XString(M)$ carries a fake-flat connection implies that the Levi-Civita connection on $\spin M$ must be flat. On the other hand, the fake-flatness condition cannot be simply dropped, for intricate reasons related to the topology of the underlying 2-bundle. A solution was achieved by \cite{Fiorenza,Sati2012} through an L$_{\infty}$-theoretical approach using adjusted Weil algebras. \cite{Kim2020,Rist2022} identified a minimal way to implement this solution on the level of strict 2-groups $\Gamma$, requiring the choice of an \quot{adjustment} for $\Gamma$. \cite{Rist2022} constructed a canonical adjustment for the string 2-group $\XString(d)$, allowing to consider \quot{adjusted} connections on principal $\XString(d)$-bundles without requiring the flatness of the manifold. The equivalence of this approach to our \cref{string-connection} was  recently established:

\begin{theorem}
{\normalfont\citep{Tellez2023}}
The equivalence of \cref{string-structures-and-lifts} between string structures and  $\XString(d)$-2-bundles lifting the structure group of $\spin M$ extends to an equivalence between geometric string structures and $\XString(d)$-2-bundles with adjusted connections. 
\end{theorem}   

We remark that geometric string structures have other descriptions that we could not include into the the present article, e.g. in the topos-theoretical approach of \cite{Schreiber2011}, used in the above mentioned work  \cite{Fiorenza,Sati2012}. 

%
%
%
%
%
%
%
%
%
%
%
%
%
%


\section{The global anomaly}

\label{bunkes-theorem}

In this section, we delve into the fundamental motivation behind geometric string structures -- the cancellation of the \quot{global} anomaly of the fermionic string.

The trajectory of a string in a spacetime $M$ is described by a smooth map $\phi: \Sigma \to M$, where $\Sigma$ is a   Riemann surface; here assumed to be closed. The fermionic path integral  can be seen as a rigorously defined section in a Pfaffian line bundle $\Pfaff(\slashed D)$ over the mapping space $C^{\infty}(\Sigma,M)$, which needs to be trivialized.
The relation between this anomaly and the first  Pontryagin class was probably first discovered rationally by \cite{Moore1984}. Notably, we have the following result.

\begin{theorem}
{\normalfont\citep{freed4}}
\label{freeds-theorem}
The first Chern class of the Pfaffian line bundle $\Pfaff(\slashed D)$ is the transgression of the first fractional Pontryagin class of $M$, i.e.,
\begin{equation*}
\mathrm{c}_1(\Pfaff(\slashed D)) = \int_{\Sigma}\ev^{*}\textstyle\frac{1}{2}\mathrm{p}_1(M)\text{,}
\end{equation*}
where $\ev: \Sigma \times C^{\infty}(\Sigma,M) \to M$ is the evaluation map. In particular, the Pfaffian line bundle is trivializable if $M$ is a string manifold. 
\end{theorem}
 
\Cref{freeds-theorem} establishes the basic relation between the anomaly of the fermionic string and string manifolds. However, in order to cancel the anomaly, a specific trivialization of the Pfaffian line bundle $\Pfaff(\slashed D)$ has to be provided. Loop-spin structures as defined in \cref{spin-structures-on-loop-space} have been assumed to provide such trivializations, just like spin structures did this in case of the fermionic particle (see \cref{introduction}), but this is -- in general -- not the case. A result of Bunke (\cref{bunke} below) shows that one needs the full information of a geometric string structure.

We set out to explain some of the details, on the basis of \cite{freed5} and \cite{bunke1}. The integrand of the problematic  path integral is the fermionic action functional 
\begin{equation}
\label{fermionic-action-functional}
S^{fer}_{\phi}(\psi) := \int_{\Sigma} \left \langle   \psi,   \slashed D_{\phi}\psi \right \rangle\; \mathrm{dvol}_{\Sigma}\text{,}
\end{equation}
where $\psi$ is a worldsheet spinor, a section in the twisted spinor bundle $S_{\phi} := S\Sigma \otimes \phi^{*}TM$ over $\Sigma$, $\slashed D_{\phi}$ is a version of the twisted Dirac operator.

The global anomaly arises when the (exponential of the) action functional \cref{fermionic-action-functional} is supposed to be integrated \quot{over all $\psi$}, which is not well-defined because the Hilbert space of sections $\mathcal{H}_{\phi}=L^2(S_{\phi})$ is infinite-dimensional and has no canonical measure. The usual way to circumvent the ill-definedness of the integral is to interpret it locally as a Berezinian integral,
whose values patch together to the canonical section of the Pfaffian line bundle $\Pfaff(\slashed D)$ of the family $\phi \mapsto \slashed D_{\phi}$.

We recall that a Berezinian integral is actually not an  integral in the sense of analysis; instead, if $V$ is a $2n$-dimensional vector space, it is the linear map
\begin{equation*}
\int^{B} : \Lambda^{*}V^{*} \to \det V^{*}
\end{equation*}
defined on homogeneous elements $\alpha \in \Lambda^kV^{*}$ by
\begin{equation*}
\int^{B}\alpha = \begin{cases} \alpha & \text{if }k=2n \\
0 & \text{else.} \\
\end{cases}
\end{equation*}
Here, $\det W$ denotes the top exterior power of a vector space $W$.
For $\alpha \in \Lambda^2V^{*}$ one has
\begin{equation}
\label{Berezinian}
\int^{B} \exp(\alpha) = \pfaff(\alpha) \text{,}
\end{equation}
the \emph{Pfaffian} of $\alpha$, a distinguished square root of the top exterior power of $\alpha$ when viewed as an element of $V^{*}\otimes V^{*}$.

The vector space $V$ to which this is applied is the sum $V_{\phi,\mu}$ of eigenspaces of the  operator $\slashed D_{\phi}^2$ below a certain spectral cut $\mu>0$. Moreover, the element $\alpha_{\phi,\mu}\in \Lambda^2V_{\phi,\mu}^{*}$ is  
\begin{equation*}
\alpha_{\phi,\mu}(\psi_1,\psi_2) :=\int_{\Sigma} \left \langle  \psi_1, \slashed D_{\phi}\psi_2 \right \rangle \;\mathrm{dvol}_{\Sigma}\text{,}
\end{equation*}  
thus designed so that the Berezinian integral \cref{Berezinian} looks like the integral over the exponential of the fermionic action functional \cref{fermionic-action-functional}, yet giving well-defined elements $\pfaff(\alpha_{\phi,\mu})\in \det V_{\phi,\mu}^{*}$.

The one-dimensional vector spaces $\det V_{\phi,\mu}^{*}$ are the fibres of a complex line bundle $\Pfaff_{\mu}(\slashed D)$ over the open set $U_{\mu} \subset C^{\infty}(\Sigma,M)$ consisting of all $\phi$ such that $\mu$ is not in the spectrum of $\slashed D_{\phi}^2$. Likewise, the assignment $\phi\mapsto \pfaff(\alpha_{\phi,\mu})$ forms a smooth section $\pfaff_{\mu}$ into $\Pfaff_{\mu}(\slashed D)$. One can then glue these locally defined line bundles to obtain a line bundle $\Pfaff(\slashed D)$ over $C^{\infty}(\Sigma,M)$, in such a way that the locally defined sections $\pfaff_{\mu}$ yield a globally defined section $\pfaff$ into $\Pfaff(\slashed D)$. The Pfaffian line bundle $\Pfaff(\slashed D)$  comes equipped with  the Quillen metric and the Bismut-Freed connection.

Summarizing, the value $\pfaff(\phi)$ of the section is regarded as a well-defined replacement of the path integral over the exponentiated fermionic action functional, paying the price that $\pfaff(\phi)$ is not a number unless a trivialization of the Pfaffian line bundle $\Pfaff(\slashed D)$ is provided. We remark that $\pfaff(\phi)$ typically has zeroes so that itself does not trivialize $\Pfaff(\slashed D)$.

Next we want to explain how a geometric string structure in the sense of \cref{string-connection} trivializes the Pfaffian bundle, thus turning the pfaffian section into a function.  For this purpose, we relate $\Pfaff(\slashed D)$ to the Chern-Simons 2-gerbe.

Analogously to the transgression functor for bundle gerbes, see \cref{transgression-functor}, there is a transgression functor for bundle 2-gerbes \citep{Waldorfa}: it takes a bundle 2-gerbe $\mathscr{C}$ with connection over $M$ to a line bundle $L\mathscr{C}$ with connection over the mapping space $C^{\infty}(\Sigma,M)$ of a closed oriented surface to $M$. It is defined similarly as for bundle gerbes: the fibre of the transgressed line bundle $L\mathscr{C}$ over $\phi:\Sigma\to M$ is composed of pairs $([\mathscr{T}],z)$ where $\mathscr{T}$ is a trivialization of $\phi^{*}\mathscr{C}$ with connection (\cref{bundle-2-gerbes-with-connection}) and $z\in \C$. The 2-Dixmier-Douady class of $\mathscr{C}$ and the first Chern class of $L\mathscr{C}$ are related by the map
\begin{equation*}
\h^4(M,\Z) \mapsto \h^2(C^{\infty}(\Sigma,M),\Z),\quad \xi \mapsto \int_{\Sigma}\ev^{*}\xi\text{.}
\end{equation*}

Now, if $\mathscr{T}$ is a geometric string structure on $M$, then the map 
\begin{equation*}
s_{\mathscr{T}}: C^{\infty}(\Sigma,M) \to L\mathscr{CS}(M),\quad \phi \mapsto ([\phi^{*}\mathscr{T}],1)
\end{equation*}
 is a smooth, nowhere vanishing section.  Taking the connection on $L\mathscr{CS}(M)$ into account, one can show that the covariant derivative of $s_{\mathscr{T}}$ is 
\begin{equation*}
\int_{\Sigma} \ev^{*}D^{\mathscr{T}}\text{,}
\end{equation*}
where $D^{\mathscr{T}} \in \Omega^3(M)$ is the covariant derivative of the string connection defined in \cref{string-connections}.
Moreover, considering the action of bundle gerbes $\mathcal{K}$ with connections on geometric string structures, we have
\begin{align}
\label{section-under-action}
s_{\mathscr{T} \otimes \mathcal{K}} = s_{\mathscr{T}} \cdot \mathrm{Hol}_{\mathcal{K}}\text{,}
\end{align}
where $\mathrm{Hol}_{\mathcal{K}}:C^{\infty}(\Sigma,M) \to \C^{\times}$ denotes  the \quot{surface holonomy} of $\mathcal{K}$ around $\phi:\Sigma \to M$. 
\begin{theorem}
{\normalfont\citep{bunke1}}
\label{bunke}
There is a canonical, isometric, connection-preserving isomorphism
\begin{equation*}
L\mathscr{CS}(M) \cong \Pfaff(\slashed D)
\end{equation*}
of line bundles over $C^{\infty}(\Sigma,M)$. 
In particular, every choice of a geometric string structure on $M$ trivializes the Pfaffian line bundle and cancels the global anomaly of the supersymmetric sigma model. 
\end{theorem}

We remark that the section $s_{\mathscr{T}}$ of $L\mathscr{CS}(M)$ depends on the \emph{geometric} part of the string structure $\mathscr{T}$, the string connection. In order to see this, consider a 2-form $\eta \in \Omega^2(M)$ and the trivial bundle gerbe $\mathcal{I}_{\eta}$ with connection $\eta$. Then, $\mathscr{T} \otimes \mathcal{I}_{\eta}$ has the same underlying string structure as $\mathscr{T}$; however, using \cref{section-under-action} we obtain
\begin{equation*}
s_{\mathscr{T} \otimes \mathcal{I}_{\eta}}(\phi) = s_{\mathscr{T}}(\phi) \cdot  \exp \left (2\pi \im \int_{\Sigma} \phi^{*}\eta \right )\text{.}
\end{equation*}

A final remark is about the role of geometric loop spin-structures, i.e. loop-spin structures (\cref{loop-spin-structure}) equipped with loop-spin connections (\cref{loop-spin-connections}), for anomaly cancellation. Apart from the transgression of bundle gerbes to line bundles on loop spaces $LM$ \cref{transgression-functor} and the transgression of bundle 2-gerbes to mapping spaces $C^{\infty}(\Sigma,M)$, it is also possible to transgress bundle 2-gerbes $\mathscr{C}$ with connection to bundle gerbes $\mathcal{G}\mathscr{C}$ with connection over $LM$, realizing geometrically the transgression homomorphism $\tau_M: \h^4(M,\Z) \to \h^3(LM,\Z)$. In fact, as a geometric counterpart of McLaughlin's theorem \cref{mclaughlins-theorem}, the transgression $\mathcal{G}\mathscr{CS}(M)$ of the Chern-Simons 2-gerbe $\mathscr{CS}(M)$ to $LM$ is canonically isomorphic to the spin lifting gerbe $\mathcal{S}_{LM}$ discussed in \cref{spin-structures-on-loop-space}, a fact that also extends to connections \citep{Waldorfb}. Via lifting theory, trivializations of $\mathcal{S}_{LM}$, in turn, correspond precisely to geometric loop-spin structures.   

We may canonically identify the double loop space of $M$ with the mapping space of the torus $\mathbb{T}=S^1 \times S^1$, i.e., $LLM = C^{\infty}(\mathbb{T},M)$. 
Then, the transgression of $\mathscr{C}$ to the line bundle $L\mathscr{C}$ over $C^{\infty}(\mathbb{T},M)$ factors through $LM$ via 
\begin{equation*}
\mathscr{C} \mapsto \mathcal{G}\mathscr{C} \mapsto L\mathcal{G}\mathscr{C}\cong L\mathscr{C}\text{.}
\end{equation*}
Under \cref{bunke}, this means that the spin lifting gerbe $\mathcal{S}_{LM}$ transgresses to the Pfaffian line bundle $\Pfaff(\slashed D)$ over $C^{\infty}(\mathbb{T},M)$: 
\begin{equation*}
L\mathcal{S}_{LM} \cong L\mathcal{G}\mathscr{CS}(M)\cong L\mathscr{CS}(M) \cong \Pfaff(\slashed D)\text{.}
\end{equation*} 
Thus, geometric loop-spin structures transgress to trivializations of $\Pfaff(\slashed D)$, and hence cancel the global anomaly only for $\Sigma=\mathbb{T}$.  As one can see from  \cref{equivalence-geometric-string-structures-loop-space}, it is precisely the addition of fusion products that allows to extend this to all surfaces.

\section{The Witten genus}

\label{Witten-genus}

In this concluding section, we explore some aspects of the relationship between string structures and the Witten genus.
The Witten genus $\mathrm{W} (M )$ of a spin manifold $M$ is  given by the power series 
\begin{equation*}
\mathrm{W}(M) := \mathrm{ind}(S(TM_{\C} -\underline{\C}^{d})) \in \Z[[q]]\text{.}
\end{equation*}
Here, $\underline{\C}^{n}$ is the trivial rank $n$ complex vector bundle over $M$, $TM_{\C}$ is the complexified tangent bundle of $M$, the difference $TM_{\C}-\underline \C^{d}$ is evaluated in topological K-theory $\mathrm{K}(M)$, the homomorphism $S: \mathrm{K}(M)\to \mathrm{K}(M)[[q]]$ sends $E\in \mathrm{K}(M)$ to 
\begin{equation*}
S(E):=\bigotimes_{l\geq 0}S_{q^{l}}(E)
,\quad
S_{q}(E) := \sum_{k\geq 0} \mathrm{Sym}^{k}E\cdot q^{k}\text{,}
\end{equation*}
where $\mathrm{Sym}^k$ denotes the $k$-th symmetric power,
and $\mathrm{ind}:\mathrm{K}(M) \to \Z$ denotes the  index of the twisted Dirac operator on $M$. We refer to \cite{stolz4} and \cite{Dessai2009} for introductions to the Witten genus.

One can check that the first coefficients of the power series $S(TM_{\C} -\underline \C^{d})$ are
\begin{flushleft}
$\underline\C^1 +(TM_{\C}-\underline \C^{d})q$
\\[0.5em]\raggedleft
$+(\mathrm{Sym}^2TM_{\C}-(n-1)TM_{\C}+\frac{n(n-3)}{2}\underline \C^1)q^2+... $
\end{flushleft}
Thus, the constant term of $\mathrm{W}(M)$ is the untwisted index, the $\widehat {\mathrm{A}}$-genus $\widehat {\mathrm{A}}(M)\in \mathrm{K}(M)$, and one may hence regard the Witten genus as a refinement of the $\widehat {\mathrm{A}}$-genus.

Heuristically, $\mathrm{W}(M)$ is the equivariant  index of a  Dirac operator acting on sections of the spinor bundle $S_{LM}$ of \cref{spinor-bundle-on-loop-space} \citep{Witten1982,Alvarez1987}. That Dirac operator has not been constructed rigorously, and could only been studied on the \emph{formal loop space}, e.g. by \cite{Taubes1989}. 

Regarding the aforementioned formulas, it's important to note that the Witten genus is independent of string structures and does not even depend on the spin structure on $M$.
However, a famous conjecture of \cite{stolz4} says that the Witten genus $\mathrm{W}(M)$ vanishes on  string manifolds that admit a metric of positive Ricci curvature.

Another relation to string structures is the following. If $M$ has dimension $d=4m$ and admits a rational string structure (see \cref{string-connections}), then  \cite{Zagier1988} proved that $\mathrm{W}(M)$ is the $q$-expansion of a modular form of weight $2m$ over $\mathrm{SL}(2,\Z)$.
The appearance of modular forms sparked the interest of homotopy theorists in the Witten genus, and indeed, the Witten genus of a string manifold can be lifted to take values in the ring $\mathrm{TMF}_{*}$ of \emph{topological modular forms} constructed by \cite{Hopkins1994}; see \cite{Douglas2014}. The ring $\mathrm{TMF}_{*}$ is the coefficient ring of the E$_{\infty}$-ring spectrum $\mathrm{TMF}$, and the Witten genus lifts to a morphism
\begin{equation*}
\sigma_\mathrm{W}: \mathrm{M}\str- \to \mathrm{TMF}
\end{equation*}
of ring spectra, the  \emph{string orientation of $\mathrm{TMF}$} \citep{Ando2010}.

We remark that the  string orientation $\sigma_\mathrm{W}$ depends on the string structure. For example, the 3-sphere $S^3$ with the canonical string structure $\mathscr{T}_{can}$ has $\sigma_\mathrm{W}=0$ whereas $\sigma_\mathrm{W}$ is non-trivial when $S^3$ is equipped with the basic string structure $\mathscr{T}_{bas}$, see \cref{string-structures}. 
In particular, an extension of the Stolz conjecture to $\sigma_\mathrm{W}$ does not hold. This situation is remarkable because it is differs from the analogy with the $\widehat {\mathrm{A}}$-genus and its corresponding homotopy-theoretical lift $\alpha:\mathrm{M}\spin - \to \mathrm{KO}$: when  $M$ has a metric of positive scalar curvature, then by Lichnerowicz' formula  $\widehat {\mathrm{A}}(M)=0$, but it is also correct  that $\alpha(M)=0$ for \emph{all} spin structures on $M$ \citep{Hitchin1974}.


%
%
%
%

%

In the remainder of this section we will review a construction  of \cite{bunke2} of a secondary invariant for $(4m-1)$-dimensional string manifolds. That invariant becomes particularly simple for $m=1$, where it takes values in $\Z_{24}$. Since $\Omega_3^{\spin -}=0$,  every  3-dimensional string manifold is the boundary of a spin manifold $Z$.
The Levi-Civita connection on $TM$ can be extended to $TZ$; moreover, by \cref{string-connections-exist} we can choose a string connection for the given string structure $\mathscr{T}$ on $M$. The Bunke-Naumann invariant is defined by
\begin{equation*}
\mathrm{BN}(M) := \int_Z {\textstyle\frac{1}{2}} \mathrm{p}_1(Z)-\int_M D^{\mathscr{T}}\in \R\text{,}
\end{equation*}
where $D^{\mathscr{T}}$ is the covariant derivative of the geometric string structure $\mathscr{T}$. 

\begin{theorem}
{\normalfont\citep{bunke2}}
The number $\mathrm{BN}(M)$ is an integer and independent of the string connection. Moreover, its reduction in $\Z_{24}$ is independent of the choice of the spin manifold $Z$, and it vanishes when $Z$ is a string manifold. Finally, it induces the isomorphism
\begin{equation*}
\mathrm{BN}: \pi_3(\mathrm{M}\str-) \to \Z_{24}.
\end{equation*}
\end{theorem}

The definition of $\mathrm{BN}(M)$ can be generalized to $4m-1$ dimensions, provided the underlying spin manifold is zero bordant, which is not automatic anymore. It takes then values in a ring $T_{2m} := \R[[q]]/(\Z[[q]] + \mathrm{MF}_{2m})$, which contains at $m=1$ the former ring $\Z_{24}$ as a subring.  \cite{bunke2} prove a secondary index theorem, the commutativity of the diagram   
\begin{equation*}
\alxydim{}{\pi_{4m-1}(\mathrm{M}\str -) \ar[r]^-{\sigma_\mathrm{W}}  & \pi_{4m-1}(\mathrm{TMF}) \ar[d] \\ \pi_{4m-1}(\mathrm{M}\str -^{\spin - = 0}) \ar@{^(->}[u] \ar[r]_-{\mathrm{BN}} & T_{2m}  }
\end{equation*}
showing that the analytical invariant $\mathrm{BN}$ computes -- in the cases where it is defined -- the homotopy-theoretic string orientation.  

\bibliography{kobib}
\end{document}